\documentclass{article}
\usepackage{arxiv}

\usepackage[utf8]{inputenc} 
\usepackage[T1]{fontenc}    

\usepackage{url}            
\usepackage{graphicx}
\usepackage{hyperref}       

\usepackage[backend=bibtex]{biblatex}
\bibliography{references}

\pdfoutput=1

\usepackage[textsize=tiny, textwidth=1cm]{todonotes}
\usepackage{subcaption}
\usepackage[group-separator={,}]{siunitx}
\DeclareGraphicsExtensions{.eps,.png}

\begin{document}

\title{Evaluating the Effect of the Financial Status to the Mobility Customs}

\author{Gerg\H{o} Pint\'{e}r\\
John von Neumann Faculty of Informatics, \\
\'{O}buda University, \\
B\'{e}csi \'{u}t 96/B, 1034 Budapest, Hungary \\
\texttt{pinter.gergo@uni-obuda.hu} \And
Imre Felde \\
John von Neumann Faculty of Informatics,\\
\'{O}buda University,\\
B\'{e}csi \'{u}t 96/B, 1034 Budapest, Hungary \\
\texttt{felde.imre@uni-obuda.hu}
}

\keywords{mobile phone data; human mobility; socioeconomic characteristics}

\maketitle{}

\begin{abstract}
    In this article, we explore the relationship between cellular phone data and housing prices in Budapest, Hungary. We determine mobility indicators from one months of Call Detail Records (CDR) data, while the property price data are used to characterize the socioeconomic status at the Capital of Hungary. First, we validated the proposed methodology by comparing the Home and Work locations estimation and the commuting patterns derived from the cellular network dataset with reports of the national mini census. We investigated the statistical relationships between mobile phone indicators, such as Radius of Gyration, the distance between Home and Work locations or the Entropy of visited cells, and measures of economic status based on housing prices. Our findings show that the mobility correlates significantly with the socioeconomic status. We performed Principal Component Analysis (PCA) on combined vectors of mobility indicators in order to characterize the dependence of mobility habits on socioeconomic status. The results of the PCA investigation showed remarkable correlation of housing prices and mobility customs.
\end{abstract}

\section{Introduction}
\label{sec:introduction}

Mobile phones and even more, smart-phones, are now fundamental parts of our life, they are practically, always with us, wherever we go, almost as if they were a part of our body. The continuous communication between a device and the Mobile Phone Networks leaves traces at the Operator's system of our mobility habits. Via these devices, the Mobile Phone Network can `sense' our movements, which is the basis of the ``Smart City'' concept \cite{batty2012smart}. The growing number of scientific works, published in the last two decades show the potential of the human mobility characterization using Call Detail Records (CDR) \cite{gonzalez2008understanding} \cite{candia2008uncovering, calabrese2011real, csaji2013exploring, pappalardo2015returners} --- among other fields --- in epidemiology, sociology and urban planning.

Human mobility analysis can work on a individuals or groups. Grouping is possible by any of the properties of the individuals. For example, using (estimated) home locations, the mobility patterns of the people who live in the same area can be compared to the dwellers of another area. Then, with using external data source like estate price of the city areas, the grouping based on the financial status of the individuals can be achieved assuming that the estate price of someone's home can properly describe their financial status.
In this paper, estate price data are combined with CDR data to evaluate the effect of the financial status to the mobility customs.

The analysis of the human movement patterns on the basis of the CDR data, that makes it possible to examine a very large population cost-effectively resulted a number of discoveries about human dynamics.
These works usually consider the population as a homogeneous group and the classification is based on some of the mobility indicators \cite{pappalardo2015returners}.
In order to extend the investigation to other characteristics of the population, the next step was adding external data sources to the Mobile Phone Network Data. Often, this external source is used to classify the data (e.g. by gender) and to analyze the classes, using the previously introduced mobility indicators.
Among others: social network data \cite{cecaj2014re}, transportation data \cite{huang2018modeling}, taxi trips \cite{furno2017fusing}, socioeconomic indicators (per capita income, education rate, unemployment rate and deprivation index) \cite{pappalardo2015using}, sale price of residential properties \cite{xu2018human}. Yes, this work also fits into the trends, using the selling prices of properties.

In this paper, we explore relations between mobile phone indicators and measures of socioeconomic status based on housing prices. In doing so, we perform statistical analysis on mobility indicators derived from a CDR dataset and information of property prices in Budapest. Assuming that housing prices characterize on a certain level the socioeconomic status we are exploring the relationship between the market price of properties at the neighborhood where residents live or work and mobility indicators. Our hypothesis is that mobile phone indicators depict different types of spatial variations for different set of housing prices and therefore hidden patterns and correlations can be revealed. The discovered interrelation between Socioeconomic Status (SES) and mobility habits of the residents may lead to a more in-depth understanding of urban life and could support the optimization public transport, or decision making for real estate developers and could gain the efficiency of retail processes.

This paper is organized as follows: Section~\ref{sec:literature_review} gives a brief summary of the relevant literature.
Section~\ref{sec:exploring_the_data}, describes the data used in this study.
Section~\ref{sec:methodology}, details the methodology used and the indicators and metrics generated. The mobility indicators obtained from the dataset recorded by cellular operator has been validated by comparing with reports of the national mini census (Section~\ref{sec:commuting}).
Then, the results of the statistical analysis have been discussed in Section~\ref{sec:results}. The summary of findings are given in Section~\ref{sec:conclusions}.

\section{Literature Review}
\label{sec:literature_review}

The understanding of human movements and recognition of behavior patterns that occur during daily life, in urban areas, requires a systematic analysis of that human mobility. The evaluation of human travel is based on observations on the individual and group levels. In the last decades, several novel datasets, based on vehicular GPS and cellular network records or social media information, became available which provided more accurate and sophisticated characterization of people's movements. Gonzales et al. used a dataset that included \num{100000} user's cellphone records, obtained for a half a year and showed that inhabitants usually visit some highly frequented locations \cite{gonzalez2008understanding}. Song et al. evaluated a dataset representing \num{50000} cellular phone user's recorded for three months. The human movements derived from the cellular information of the dataset has been found to be highly predictable \cite{song2010modelling}. Numerical indicators (i.e. entropy-based metrics, radius of gyration) have been calculated to quantify the temporospatial distribution of people and their movements. Parija et al. \cite{parija2019mobility} applied Profile-based paging algorithm on CDRs in order to discover personal human mobility patterns. The results suggest that the proposed algorithm is three times more efficient than conventional paging and two times more effective than various other intelligent paging algorithms. These researches showed the hidden characteristics of individual movements and gave a tangible boost for further scientific efforts.

The demographic metrics and Socioeconomic Status (SES) seems the have significant relationship to individual travel behavior. Early studies aimed to investigate correlation of the human travel characteristics and of SES \cite{hanson1981travel, kwan1999gender}. Cottineau and Vanhoof \cite{cottineau2019mobile} developed a model to explore the relationship between mobile phone data and traditional socioeconomic information from the national census in French cities. Mobile phone indicators estimated from six months Call Detail Records, while census and administrative data are used to characterize the socioeconomic organization of French cities. The findings show that some mobile phone indicators relate significantly with different socioeconomic organization of cities. Pokhriyal et al. \cite{pokhriyal2017combining} used a computational framework to accurately predict the Global Multidimensional Poverty Index (MPI) in Senegal based on  environmental data and CDR. The methodology provides the accurate prediction of important dimensions of poverty: health, education, and standard of living (the estimations have been validated using deprivations calculated from census).

In the last decade, CDRs have become a standard information source for analyzing the social characteristics of human mobility. A research work \cite{xu2015understanding} that investigated the people's daily activities in Shenzhen, China. The results identified a so-called, ``north--south'' differences of human activity, which findings are in good agreement with the socioeconomic divide in the city.

Some investigations suggest that the mobile phone data can be used to predict individual SES \cite{blumenstock2015predicting} or regional socioeconomic characteristics \cite{vscepanovic2015mobile}. Xu et al. \cite{xu2018human} used an analytical framework on large scale mobile phone and urban socioeconomic datasets to evaluate mobility patterns and SES. Six mobility indicators, housing prices and per capita income in Singapore and Boston have been used to analyze the socioeconomic classes. It was found that phone users who are generally wealthier, tend to travel shorter distances in Singapore, but longer, in Boston. The research brought interesting findings, but also showed that the relationship between mobility and socioeconomic status is worth investigation in other cities and countries as well.

Castillo et al. calculated Human Development Index for locally available data for Ecuador to describe socioeconomic status and used in comparison to their mobile phone based approach \cite{castillo2018silence}.
Barboda et al. found significant differences in the average travel distance between the low and high income groups in Brazil \cite{barbosa2020uncovering}.
This study fits into the trend, focusing on a Central European capital, Budapest.

Epidemiology is used to be mentioned as a potential application of human mobility studies, with some applications like \cite{brdar2016unveiling}, but the COVID-19 pandemic prioritised the connection between the cellular phone based mobility analytics and the epidemiology.
The term `Digital Epidemiology' can used when working with data that was not generated with the primary purpose of epidemiological studies \cite{salathe2018digital}.

Willberg et al. identified a significant decrease of the population presence in the largest cities of Finland after the lockdown compared to a usual week \cite{willberg2021escaping}.
Bushman et al. analyzed the compliance to social distancing in the US using mobile phone data \cite{bushman2020effectiveness}. Gao et al. found negative correlation in stay-at-home distancing and COVID-19 increase rate \cite{gao2020association}.
W. D. Lee et al. examined the effect of the SES to the mobility changes during the lockdown, and found that the mobility of the wealthier subscribers decreased more significantly in England\cite{do2021association}.

The data, used in this study, predate the COVID-19 pandemic and solely focuses the `normal' life of Budapest, and evaluates the effect of the financial status to the mobility customs.

\section{Exploring the Data}
\label{sec:exploring_the_data}

The data set used in this study is provided by Vodafone Hungary.
The CDR data were collected from Budapest, Capital of Hungary and the surrounding county. Vodafone Hungary is one of the three mobile phone operators providing services in Hungary. The market share of the three big operators in Hungary has not changed significantly in the last few years. Vodafone Hungary had 25.5\% in 2017 Q2 nationwide \cite{nmhh_mobile_market_report}.

The communication between a cellular device and the mobile phone network can be divided into two categories: (i) An administrative communication maintaining the connection with the service, for example, registration of the cell-switching that can be called passive communication. (ii) When the device actively uses the network for voice calls, message or data transfer, that can be called active communication.
The available data contains only the active communication, which is sparser, so it cannot be used to track continuous movements.

The raw CDR data contains some kind of hash tag to identify the SIM (Subscriber Identity Module), a timestamp that is truncated to 10 seconds, and an ID of the cell, thus, a subscriber can be mapped to a geographic location in a given time. These are extended with the type of the customer (business, consumer), the type of the subscription (prepaid, postpaid), the age and gender of the subscriber and the Type Allocation Code (TAC) of the device. The TAC is the first eight digits of the IMEI (International Mobile Equipment Identity) number that refers to the manufacturer and the model of the device wherein the SIM card is active.
These values are also present in every record, so for example the device changes can be tracked as well.

As for the cells, a separate table was provided with a cell ID, the geographic location of the cell centroid, the area of the cell and the distance between the centroid and the base station. These values are an estimation based on a momentary state, especially with the UMTS cells due to their breathing mechanism, that can change the geographic size of the serving area for load balancing. The heavily loaded cells shrink and the neighboring ones grow to compensate \cite{al2003performance}.

The rationale of this `wide' format may be that the subscriber data and the device can be changed within the observation period. This occurs in about 3000 times, during the observation period.
The owner of the subscription can change its details and of course change the device if they bought a new mobile phone for example.
Subscriber and customer type was provided for every SIM, but age and gender was missing in many cases, presumably due to the privacy options requested by the subscriber.

The records do not include neither the type of the activity (voice call, message, data transfer) nor the direction (incoming, outgoing) and there was no data provided by the operator to resolve the TACs to manufacturer and model.

\begin{figure}[ht]
    \centering
    \includegraphics[width=\linewidth]{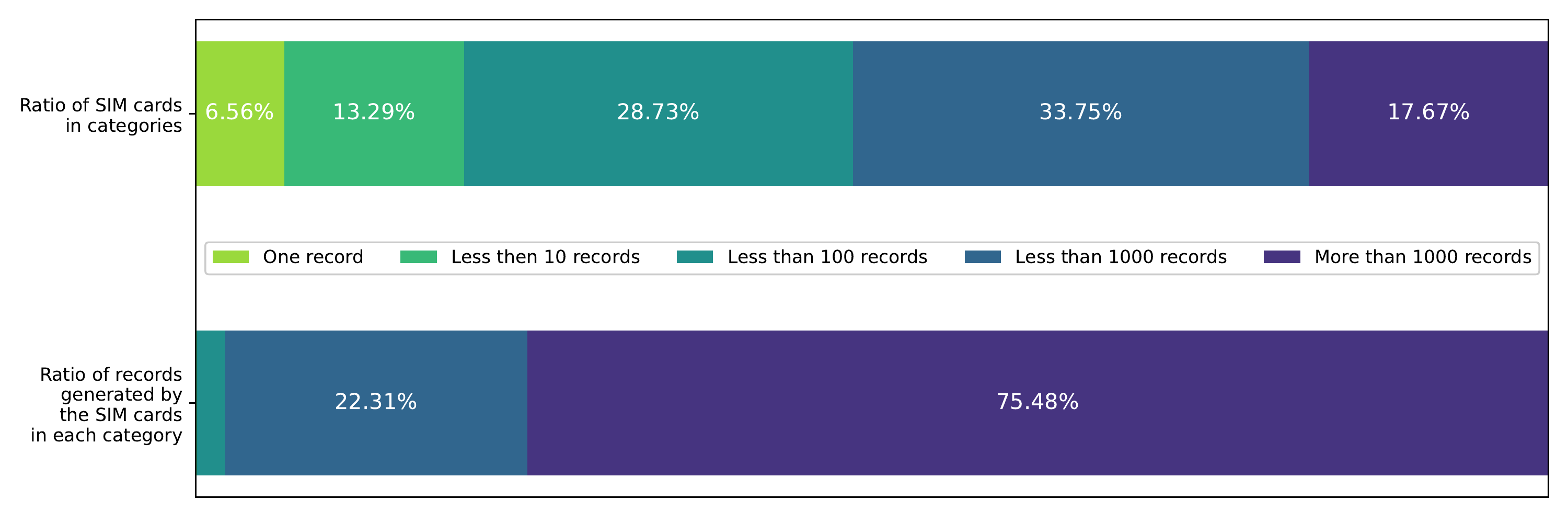}
    \caption{SIM cards in the 2017-04 data set categorized by the number of activity records. The categories are: only 1 record, 1 to 10 records, 10 to 100 records, 100 to 1000 records and greater than 1000 records. The SIM cards in the last category (17.7\% of the SIM cards) provide the majority (75.48\%) of the activity.}
    \label{fig:vod201704_activity_categories}
\end{figure}

\begin{figure}[ht]
    \centering
    \includegraphics[width=\linewidth]{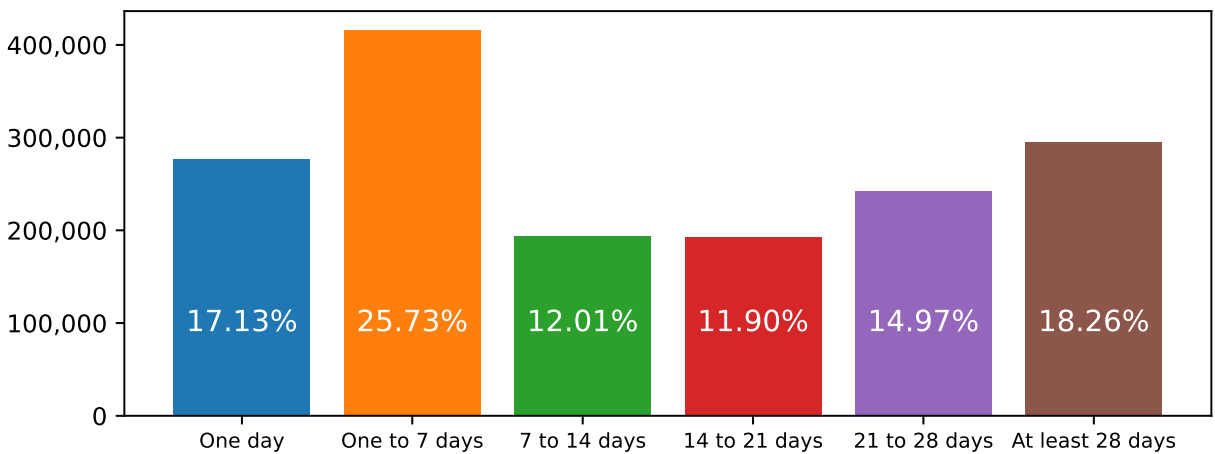}
    \caption{SIM card distribution in the 2017-04 data set by the number of active days.
    }
    \label{fig:vod201704_activity_by_days}
\end{figure}

The `April, 2017' CDR data set includes mobile phone network activity of the Vodafone users from Budapest (and the surrounding areas) in April, 2017. This contains \num{955035169} activity records, from \num{1629275} SIM cards.
Figure~\ref{fig:vod201704_activity_categories}, shows the activity distribution between the activity categories of the SIM cards. Only 17.67\% of all the SIM cards, that have more than 1000 activity records, provide the majority (75.48\%) of the mobile phone activity during the observation period.
Figure~\ref{fig:vod201704_activity_by_days}, shows the distribution of the SIM cards by the number of active days. Only about one-third (33.23\%) of the SIM cards have activity on at least 21 different days.
Despite of the relatively large number of SIM cards, present in the data set, most of them are not active enough to provide enough information about their mobility habits. For the exact selection criteria, see Section~\ref{sec:selecting_active_sims}.

\subsection{Estate Price Data}
\label{sec:estate_price}

Property estate price data are provided by the \textit{ingatlan.com} estate selling website. The data contains slightly more than 60 thousand estate locations, floor spaces and selling prices from the advertisements. The prices may not be the actual value that the buyer paid, but even if there was some bargaining, the order of magnitude should be reasonably accurate.

The data are from 2018, not from the same year as the CDR data, but the price differences between the areas of Budapest have not changed significantly during the last few years, so it should be adequate to describe the average estate price of an area.
To do so, the price of one square meter, is calculated from the floor space and the selling price. In this way, the price-level of two different estates in two very different parts of the city can be compared.

\section{Methodology}
\label{sec:methodology}

\begin{figure}[ht]
    \centering
    \includegraphics[width=\linewidth]{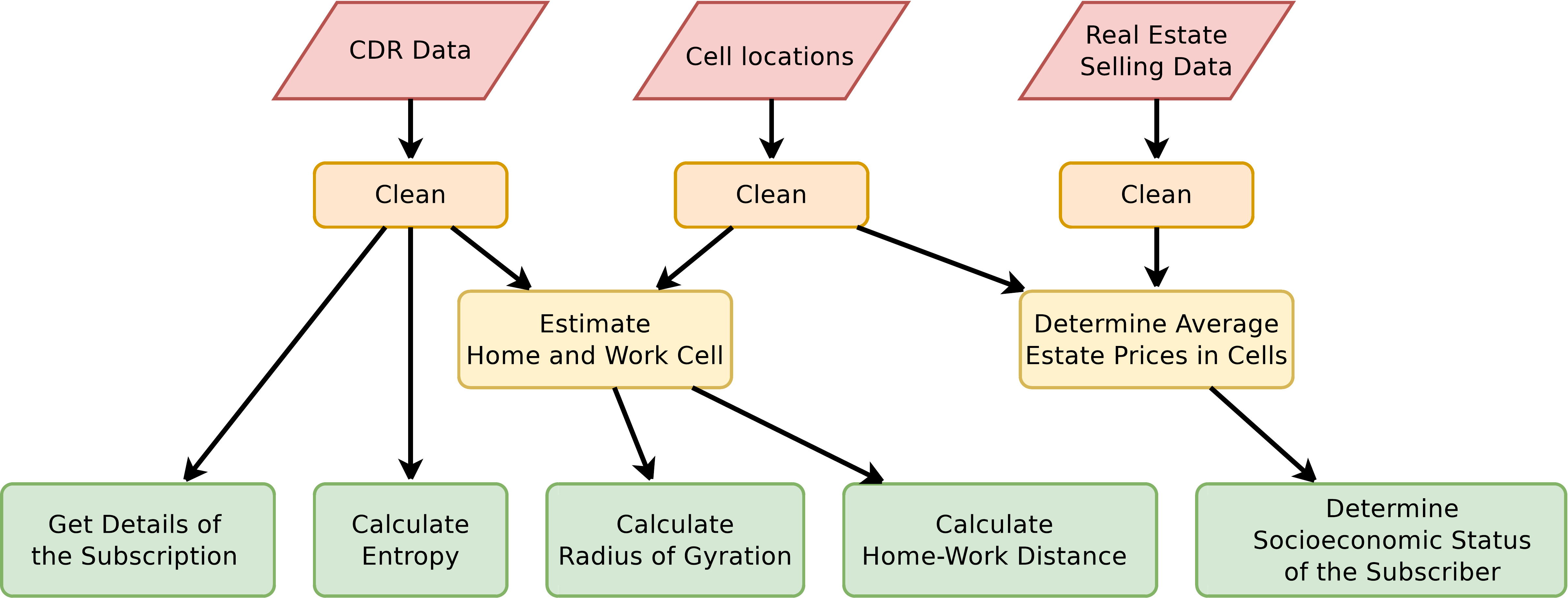}
    \caption{The simplified processing the Mobile Phone data along with the Cell information and the Estate Price data used to determine the financial status of the subscribers.}
    \label{fig:process}
\end{figure}

A computational framework has been developed to process the CDR data, including a preprocessing with cleaning module and modules  dedicated to the Home and Work locations or calculate the mobility indicators (like Radius of Gyration and Entropy). This simplified process can be seen in Figure~\ref{fig:process}.

\subsection{Data Preparation}
\label{sec:data_preparation}

As the received data had a wide format, they were normalized before importing into the database. The CDR table only contains the SIM ID, the timestamp and the cell ID. a table has been introduced to store the SIM related properties (like subscription type, costumer type, age, gender) and another to store the cell properties (cell centroid and base station coordinates).

The subscriber type and subscription type is provided for all of the SIM cards, whereas age, gender is frequently missing.
For a group of SIM cards, some of these values has changed during the data set. These changes could be properly realistic as the subscription type changes, the subscriber (the ownership) or the subscription plan is changed and very often the change is something like a known age becomes unknown or vice versa. This affects about 2000 devices out of about 1.6 million.
In these cases, the subscriber properties were set to unknown without, trying to decide which value should be used.

\subsubsection{Cell-Map Mapping}
\label{sec:cellmap_mapping}

Most of the cases only the base stations' coordinates are provided by the operator, therefore, it is common practice \cite{pappalardo2016analytical,csaji2013exploring,vanhoof2018comparing,candia2008uncovering,novovic2020uncovering} at CDR processing to use the base stations to map Call Detail Records to geographic locations and use Voronoi tessellation to estimate the covered area of the base station.

Vodafone Hungary provided not only the base station coordinates, but the (estimated) cell centroids for the cells. With this, the position of the SIM cards is known with a finer granularity over those with only the base station locations. This is because a base station can serve several cells.

The close cells within 100 m are merged using the DBSCAN algorithm of the Scikit-learn \cite{scikit-learn} Python package using the cell activity as weight and the Voronoi tessellation is applied for merged cell centroids, similarly as in  \cite{fiadino2017call}.
The estate price data (see Section \ref{sec:estate_price}) are mapped to cells via these voronoi polygons.

\subsection{Mobility Indicators}
\label{sec:mobility_indicators}

There are some indicators that widely used in the literature to characterize human mobility, like Radius of Gyration, Entropy or the distance between Home and the Work locations (Section~\ref{sec:work_home}). These indicators are determined for every subscribers.

\subsubsection{Radius of Gyration}
\label{sec:gyration}

The Radius of Gyration \cite{gonzalez2008understanding} defines a circle, where an individual can usually be found.
It was originally defined in Equation~(\ref{eq:gyration}), where $L$ is the set of locations visited by the individual, $r_{cm}$ is the center of mass of these locations, $n_i$ is the number of visits or the time spent at the i-th location.
\begin{equation}
    \label{eq:gyration}
    r_g = \sqrt{\frac{1}{N} \sum_{i \in L}{n_i (r_i - r_{cm})^2}}
\end{equation}

\subsubsection{K-Radius of Gyration}
\label{sec:k_gyration}

The K-radius of Gyration is calculated using only the $k$ most frequent locations of the individual \cite{pappalardo2015returners}, defined by Equation~(\ref{eq:k_gyration}). Pappalardo et al, used this approach to classify individuals by their mobility customs. Two classes named ``Returners'' and ``Explorers'' have been defined by the value of gyration. Returners are who spend most of their time between the $k$ most frequent locations ($r_g^{(2)} > r_g / 2$), in contrast to the explorers whose activity area cannot be described with the $k$ most frequent locations ($r_g^{(2)} < r_g / 2$) \cite{xu2018human}.
\begin{equation}
    \label{eq:k_gyration}
    r_g^{(k)} = \sqrt{\frac{1}{N_k} \sum_{i = 1}^{k}{n_i (r_i - r_{cm}^{(k)})^2}}
\end{equation}

\subsubsection{Entropy}
\label{sec:entropy}

The entropy of the visited locations characterize the diversity of the individual's movements, defined as Equation~(\ref{eq:entropy}), where $L$ is the set of locations visited by the individual, $l$ represents a single location, $p(l)$ is the probability of an individual being active at a location $l$ and $N$ is the total number of activities for an individual \cite{pappalardo2016analytical,cottineau2019mobile}.
\begin{equation}
    \label{eq:entropy}
    e = - \frac{\sum_{l \in L}{p(l) \log p}}{\log N}
\end{equation}

The generalization of the Entropy is the term, Travel Diversity \cite{xu2018human}. Instead of the diversity of the locations it determines the diversity of the travels between consecutive locations. It can be calculated to $k$-length transitions, where $k=1$ gives back the location entropy. It is possible to consider the transitions with or without direction, contingent on whether the difference between $L_1 \rightarrow L_2$ and $L_2 \rightarrow L_1$ is important or not.

\subsection{Evaluating Work and Home Locations}
\label{sec:work_home}

Most of the inhabitants in cities are spending significant time of a day at two locations: their homes and work places. In order to find the relationship between these most important locations and SES, first their positions of these locations have to be determined.
There are a few approaches used to find home locations via mobile phone data analysis  \cite{ahas2010using,bojic2015choosing,xu2015understanding}.

The work location is determined as the most frequent cell where a device is present during working hours, on workdays. Working hours are considered from 09:00 to 16:00.
The home location is calculated as the most frequent cell where a device is present during the evening and the night on workdays (from 22:00 to 06:00) and all day on holidays. Although people do not always stay at home on the weekends, it is assumed that most of the activity is still generated from their home locations.

\begin{figure}[ht]
    \centering
    \includegraphics[width=\linewidth]{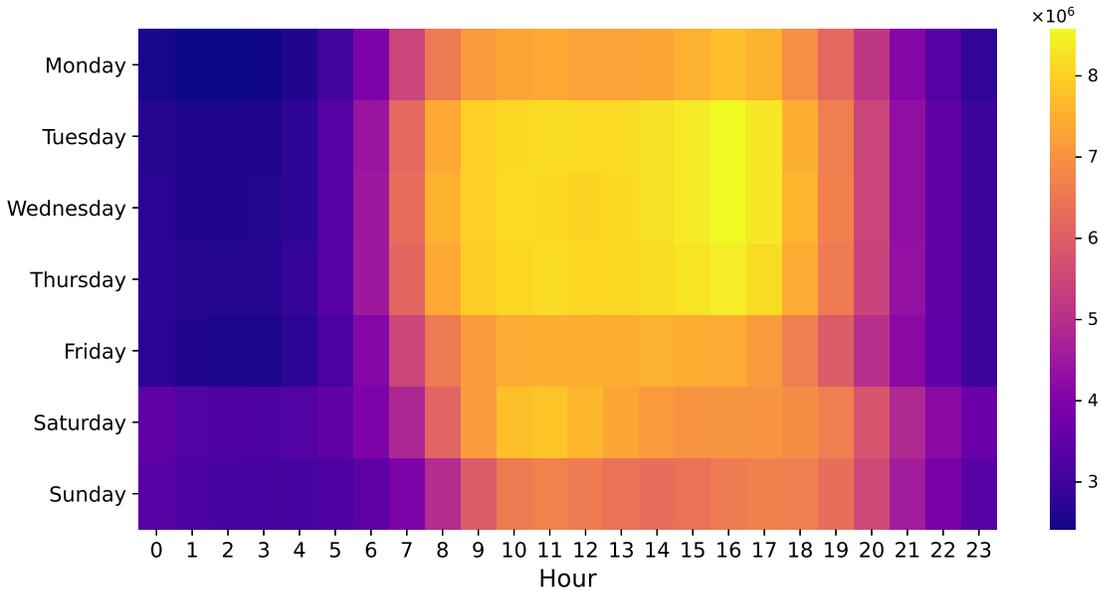}
    \caption{The mobile phone activity distribution by days of week and hours based on the 2017 data set. Mondays and Fridays are not so bright as the other workdays, probably because of Easter Monday and Good Friday that are holidays in Hungary.}
    \label{fig:vod201704_day_hour_activity}
\end{figure}

Most of the mobile phone activity occurs in the daytime (see Figure~\ref{fig:vod201704_day_hour_activity}), which is associated with work activities. This may cause the home location determination inaccurate or even impossible for some devices.

This method assumes that everyone works during daytime and rests in the evening. Although in 2017, 6.2\% of the employed persons worked regularly at night in Hungary \cite{eurostat_night_workers}, the current version of the algorithm does not try to deal with the night-workers. Some of them might be identified as a regular worker but their work and home locations are mixed.

\subsection{Selecting Active SIMs}
\label{sec:selecting_active_sims}

As showed in Section~\ref{sec:exploring_the_data}, most of the SIM cards do not have enough activity to provide adequate information about the subscribers mobility customs. Thus, only those SIMs are considered active enough, since that represents activity for at least 20 days, the daily mean on weekdays, is at least 40 activity records and at least 20 on weekends. Additionally, the average activity of the SIM cannot be more the 1000 to filter out the SIM card that possibly do not operate in a cell phone, but in a 3G modem for example.

Note that activity can be either voice call, text message or data transfer, also both incoming or outgoing, hence, they cannot be distinguished in the dataset.

\section{Analysis of Commuting}
\label{sec:commuting}

In order to verify the reliability and accuracy of the method proposed for the home and work location estimation, a comparative study is performed on the mobility indicators and the information processed from the mini census obtained in Hungary.

In Hungary, a census is obtained every 10 years and a micro-census with a 10\% corpus at the halftime. The last census was performed in 2011, while the last micro-census was in 2016. Based on these surveys, commuting to Budapest is analyzed in studies like \cite{miklos2016munkaero,koltai2020ingazas}. These studies are used as the reference for comparing our results for commuting.

\begin{figure}[t!]
    \centering
    \begin{subfigure}[t]{0.485\linewidth}
        \centering
        \includegraphics[width=\linewidth]{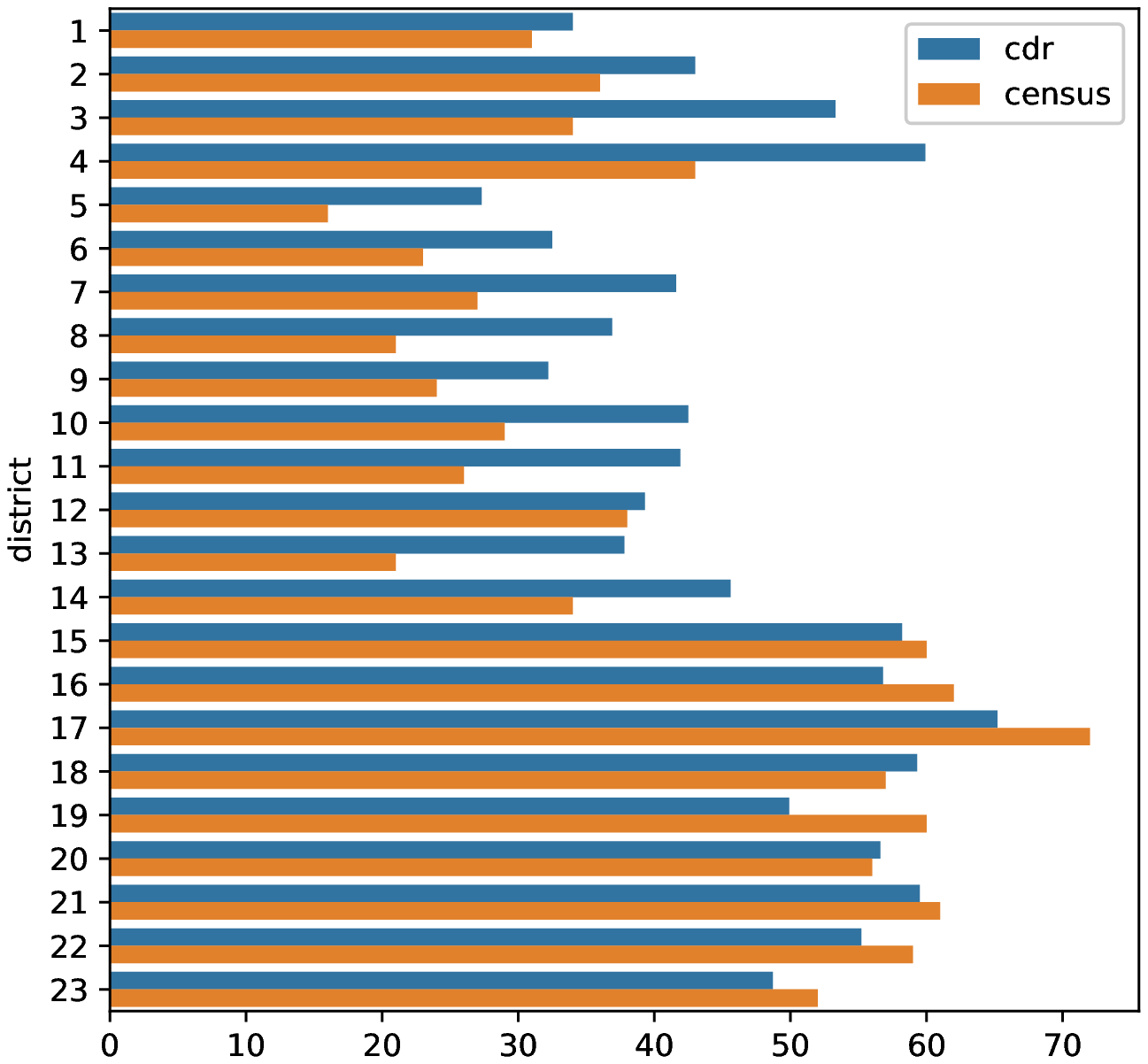}
        \caption{Commuters from the district (\%)}
        \label{fig:commuting_combined_district}
    \end{subfigure}
    ~
    \begin{subfigure}[t]{0.485\linewidth}
        \centering
        \includegraphics[width=\linewidth]{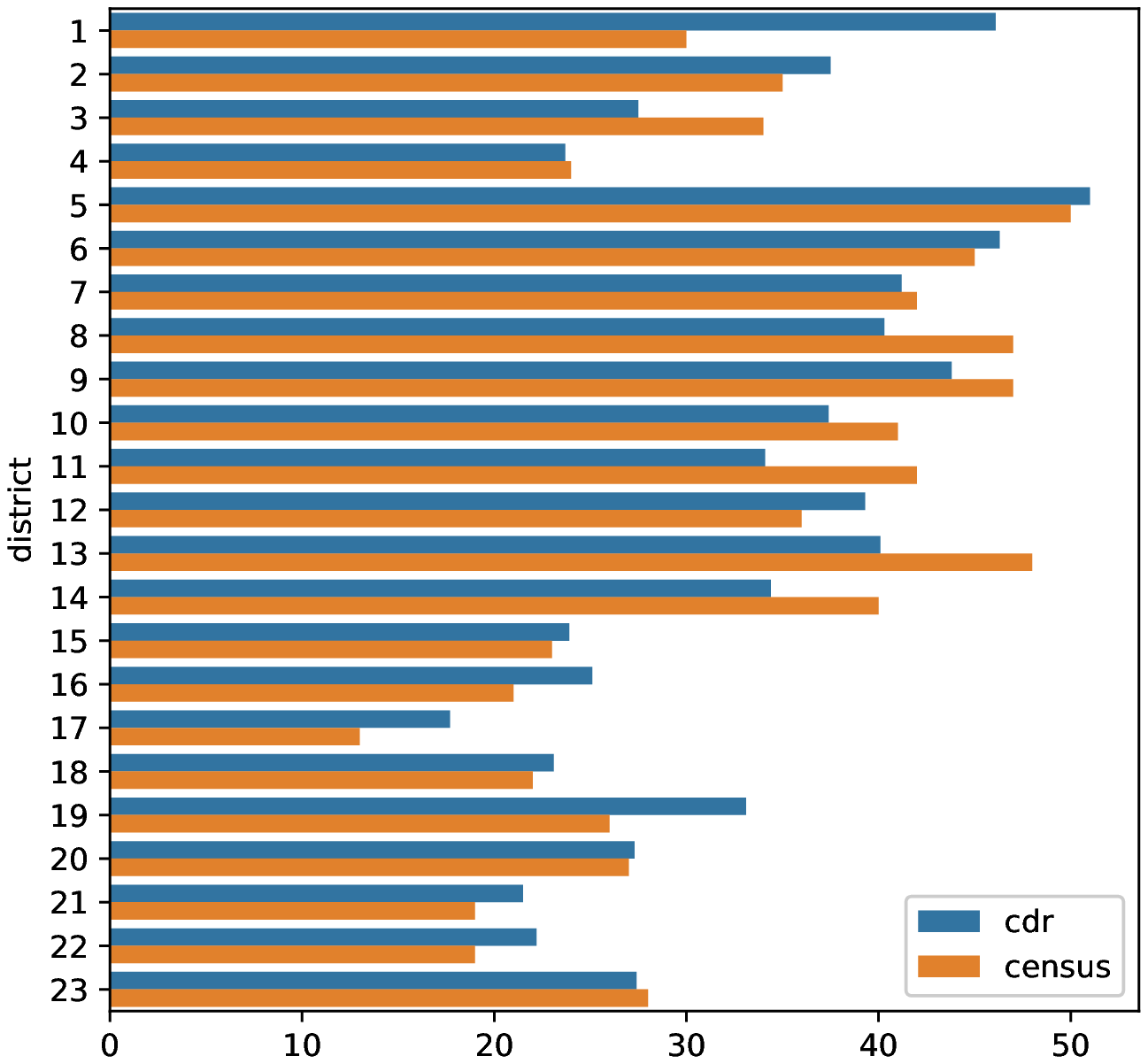}
        \caption{Commuters from Budapest (\%)}
        \label{fig:commuting_combined_budapest}
    \end{subfigure}
    \vfill
    \vspace{.75em}
    \begin{subfigure}[t]{0.485\linewidth}
        \centering
        \includegraphics[width=\linewidth]{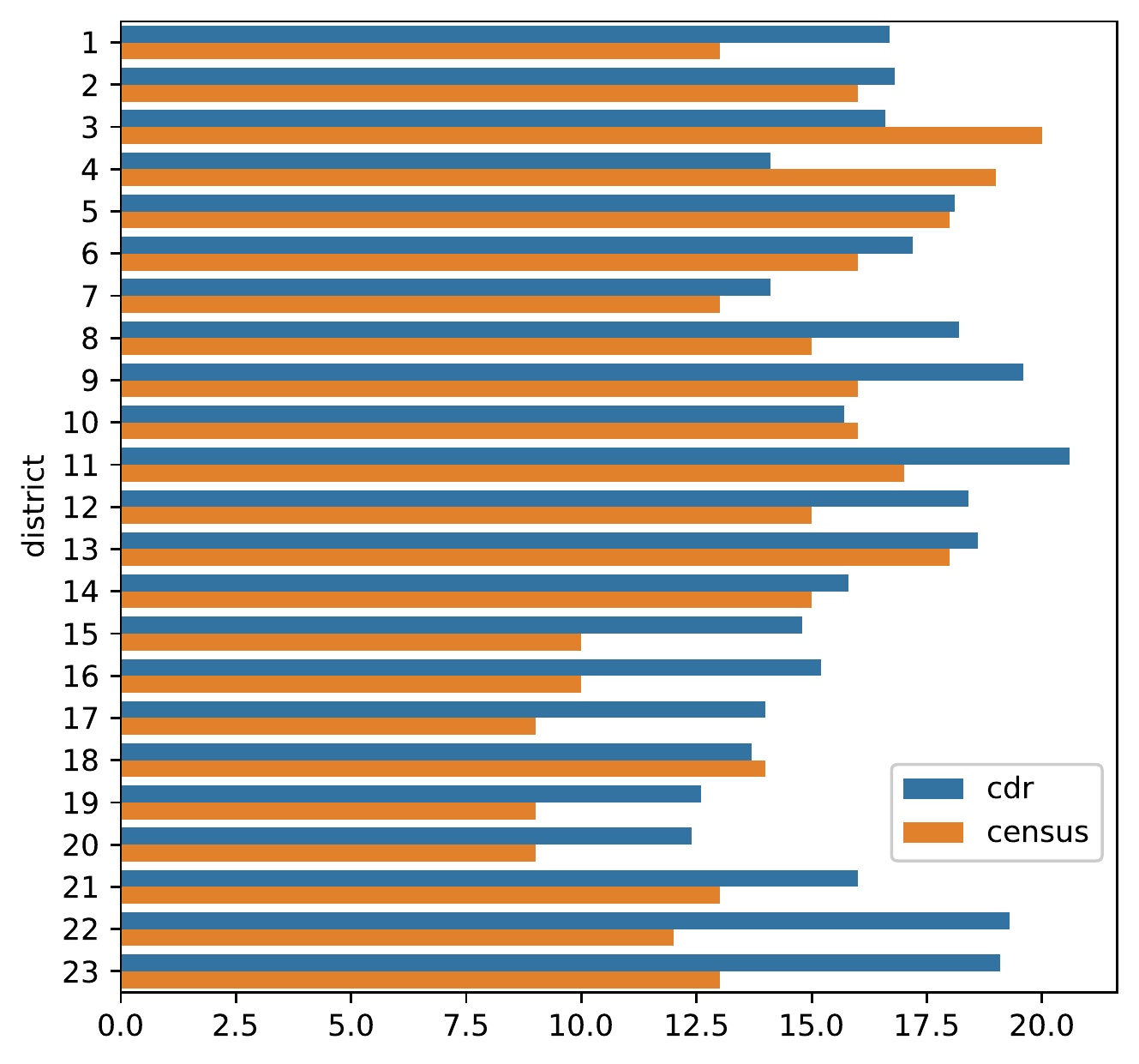}
        \caption{Commuters from the agglomeration (\%)}
        \label{fig:commuting_combined_agglomeration}
    \end{subfigure}
    ~
    \begin{subfigure}[t]{0.485\linewidth}
        \centering
        \includegraphics[width=\linewidth]{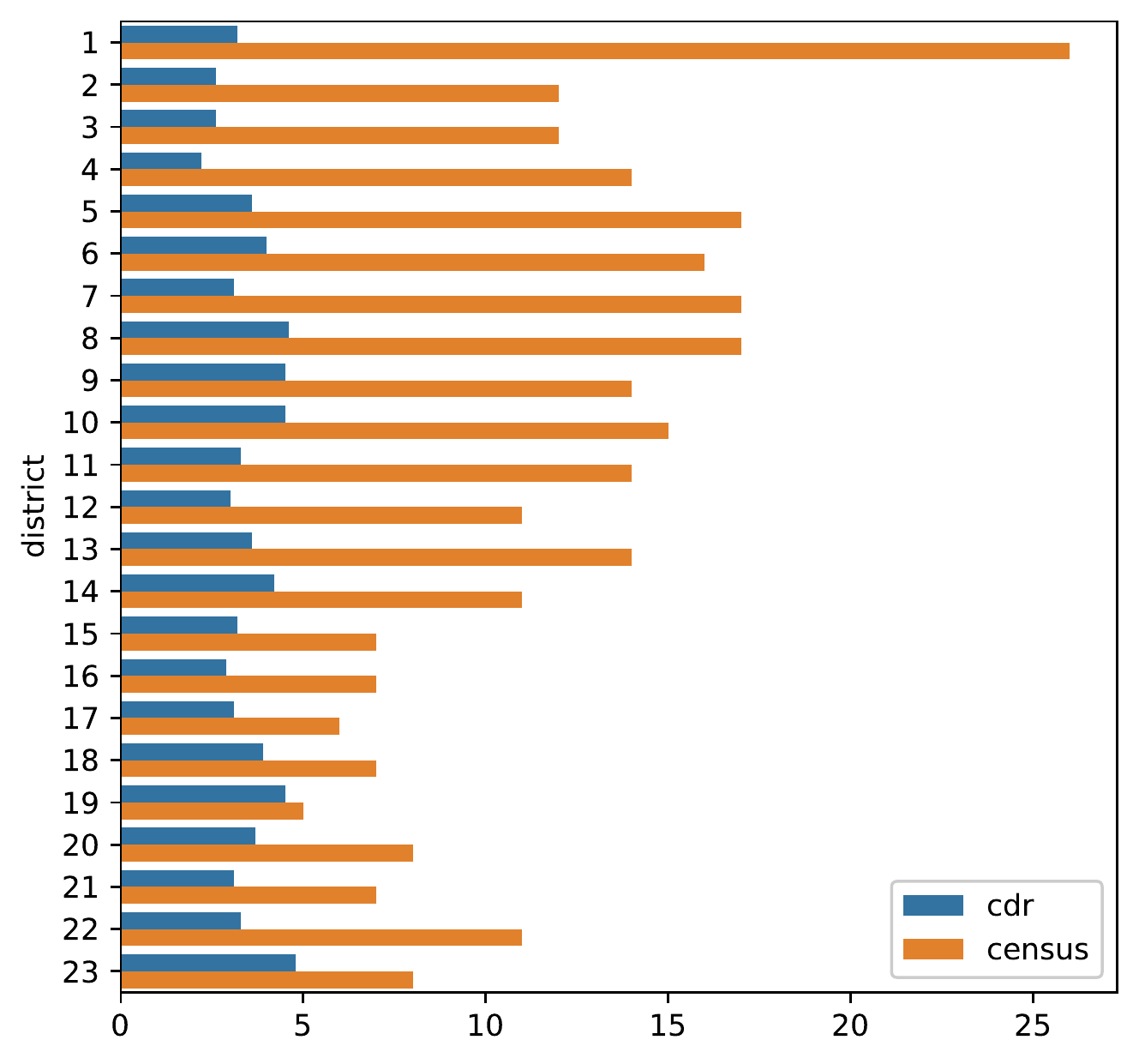}
        \caption{Commuters from outside the agglomeration (\%)}
        \label{fig:commuting_combined_other}
    \end{subfigure}
    \caption{Comparision between the CDR and the census based \cite[Figure 1]{koltai2020ingazas} commuting ratios}
    \label{fig:commuting_combined}
\end{figure}

Figure~\ref{fig:commuting_combined}, shows the comparison between the CDR and the census based\cite[Figure 1]{koltai2020ingazas} travelling ratios of the commuters by the districts of Budapest and the home location category. People who work in Budapest are represented, and the home location can be i) the same district where one works (Figure~\ref{fig:commuting_combined_district}), ii) another district of Budapest (Figure~\ref{fig:commuting_combined_budapest}), iii) the agglomeration (Figure~\ref{fig:commuting_combined_agglomeration}) and iv), other settlements outside the agglomeration (Figure~\ref{fig:commuting_combined_other}).

Good agreement mostly within Budapest has been found on the proportions of the commuters. The most significant difference can be seen with the ``outside agglomeration'' category(Figure~\ref{fig:commuting_combined_other}). This deviation is however originated to the content of the data source, as the mobile phone network data, used in this study, covers mainly the area of Budapest. It also contains phone activities from the surrounding county, but by moving away from Budapest, the available data decreases.

The fraction of workers who have their homes in the same district are very close to the census data in the outer districts (15-23), but generally overestimated in the core districts (1, 5-9) and the inner districts (2-4, 10-14). The workers from other districts group (Figure~\ref{fig:commuting_combined_budapest}) shows the best match to the census data (where the CDR should have the best quality), while the agglomeration is somewhat overestimated in many districts.

\begin{figure}[t!]
    \centering
    \begin{subfigure}[t]{0.485\linewidth}
        \centering
        \includegraphics[width=\linewidth]{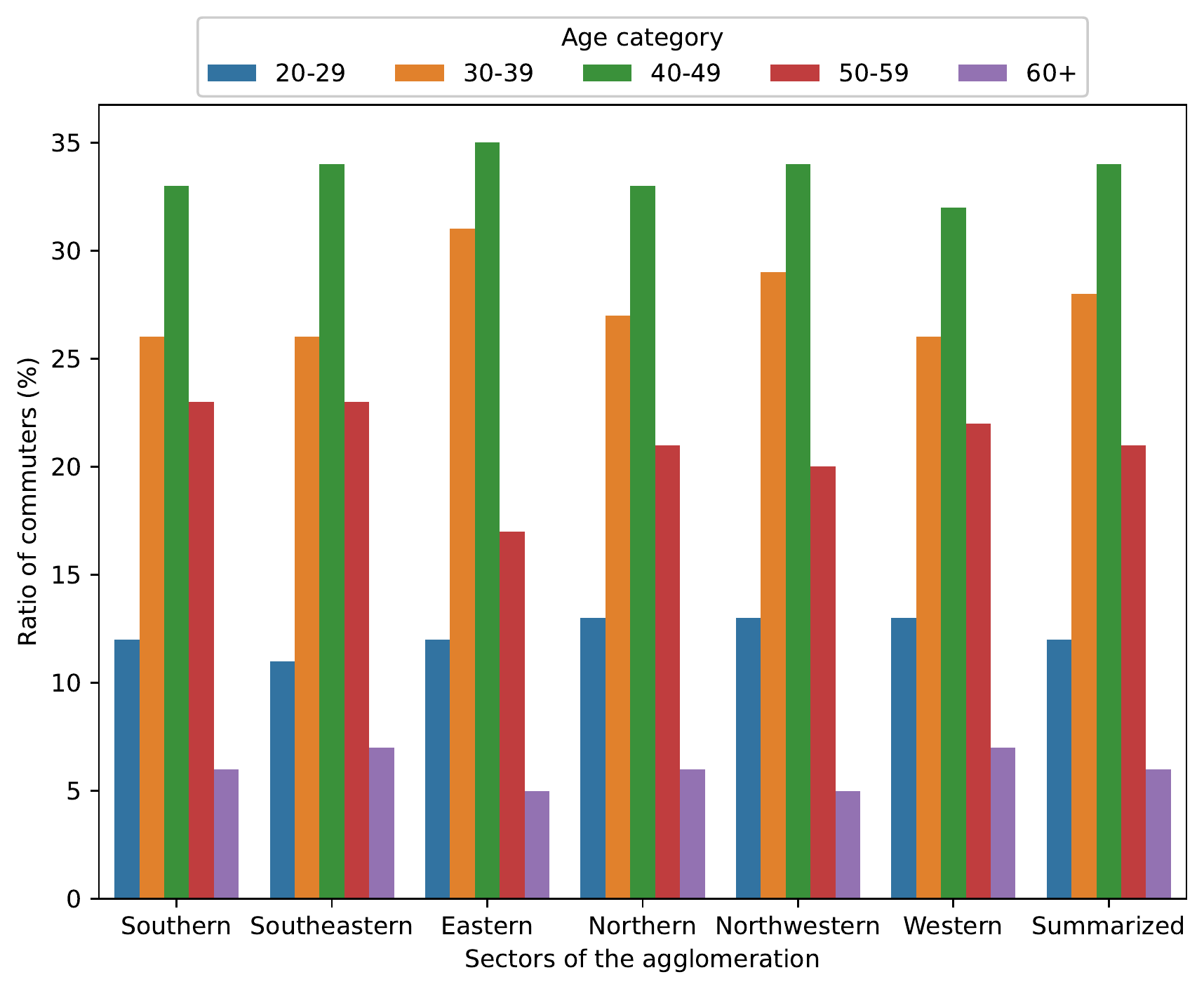}
        \caption{Distribution of the commuters by age categories and the sectors of the agglomeration (\%), data is taken from \cite[Table 1]{koltai2020ingazas}}
        \label{fig:r2t1_census}
    \end{subfigure}
    ~
    \begin{subfigure}[t]{0.485\linewidth}
        \centering
        \includegraphics[width=\linewidth]{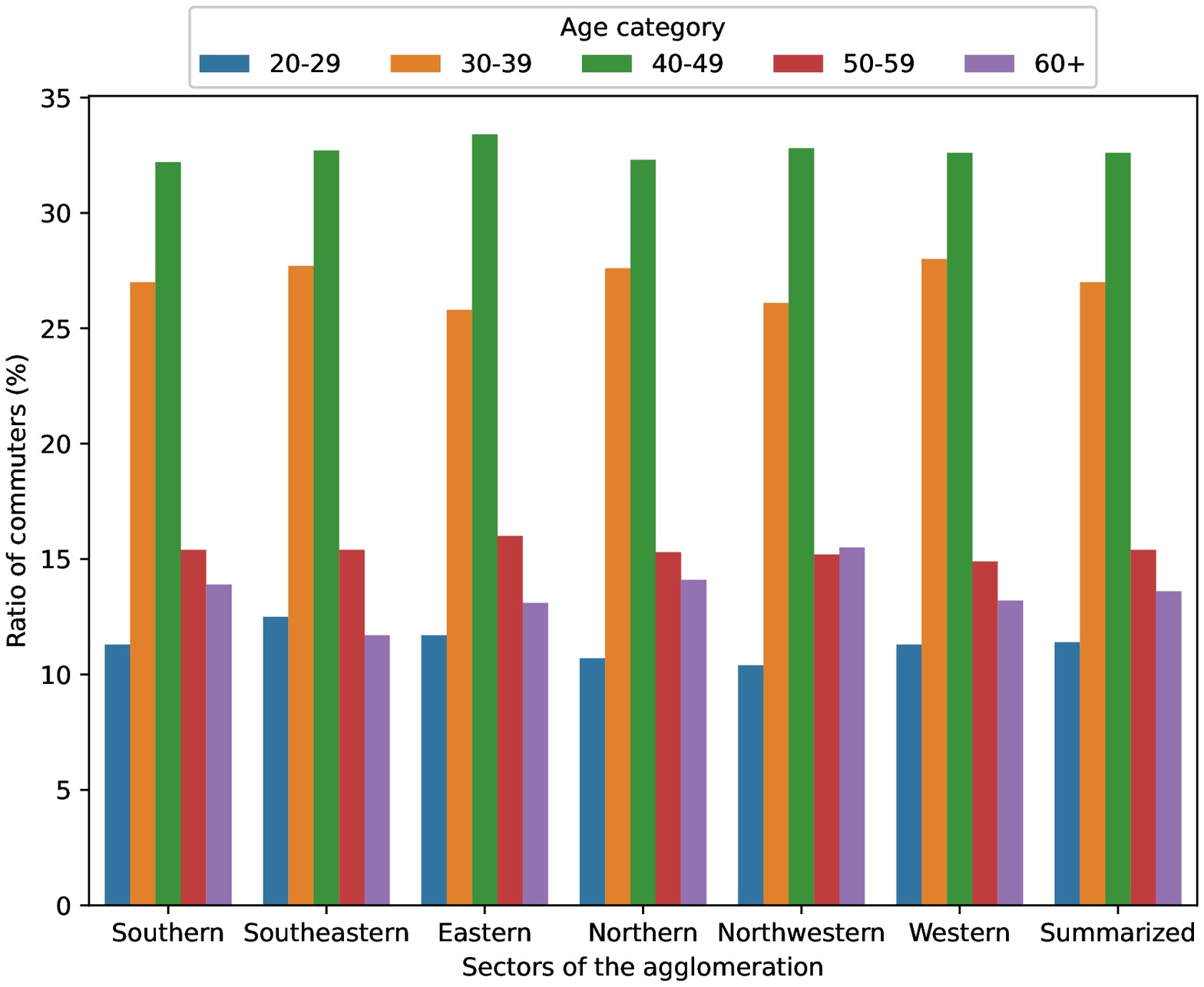}
        \caption{Distribution of the commuters by age categories and the sectors of the agglomeration (\%), based on the CDR}
        \label{fig:r2t1_cdr}
    \end{subfigure}
    \caption{Comparision of commuting to Budapest from the sectors of the agglomeration by age categories}
\end{figure}

In \cite{miklos2016munkaero}, there are more detailed analysis in regard of the commuting from the agglomeration, that is divided to six sectors and the commuting is examined by origin (home sector, occasionally by towns) and destination (district group of Budapest). Figure~\ref{fig:r2t1_census}, shows the distribution of the commuters by age categories and the sector as the home location. Only those commuters are examined who work in Budapest.

It is difficult to note from the paper, that what is the upper limit of the over 60 age category. The people who usually go to work, are assumed to be younger than 65 years old (as the current retirement age in Hungary), although people can work over 65.
In the CDR based figure (Figure~\ref{fig:r2t1_cdr}), the 60+ means over 60 and less than 100. However there are not many subscribers over 70 , only the 1.87\% of the active SIM card (Section~\ref{sec:selecting_active_sims}) are owned by people older than 70 years.
Furthermore, it has to be noted that the age information belongs to the owner of the subscription, not necessarily to the actual user of the phone.

Comparing data obtained by the mini census and the cellular information, acceptable agreements have been found on the trends and measures of the distribution of the commuters by age categories. The most significant difference between the census and the CDR based data are within the `60+' and the `50--59' categories. The former could be the result of the interpretation of the upper limit. This could also affect the other categories, especially the `50--59' category. The `20--29', `30--29' and `40--49' categories shows very similar results with both of the approaches. Based on the similarity of mobility measures given by the census and the CDR processing, the methodology proposed seems to be reliable to apply for determination of mobility indicators.

\section{Data Preparation for Statistical Analysis}

\begin{figure}[t!]
    \centering
    \begin{subfigure}[t]{0.485\linewidth}
        \centering
        \includegraphics[width=\linewidth]{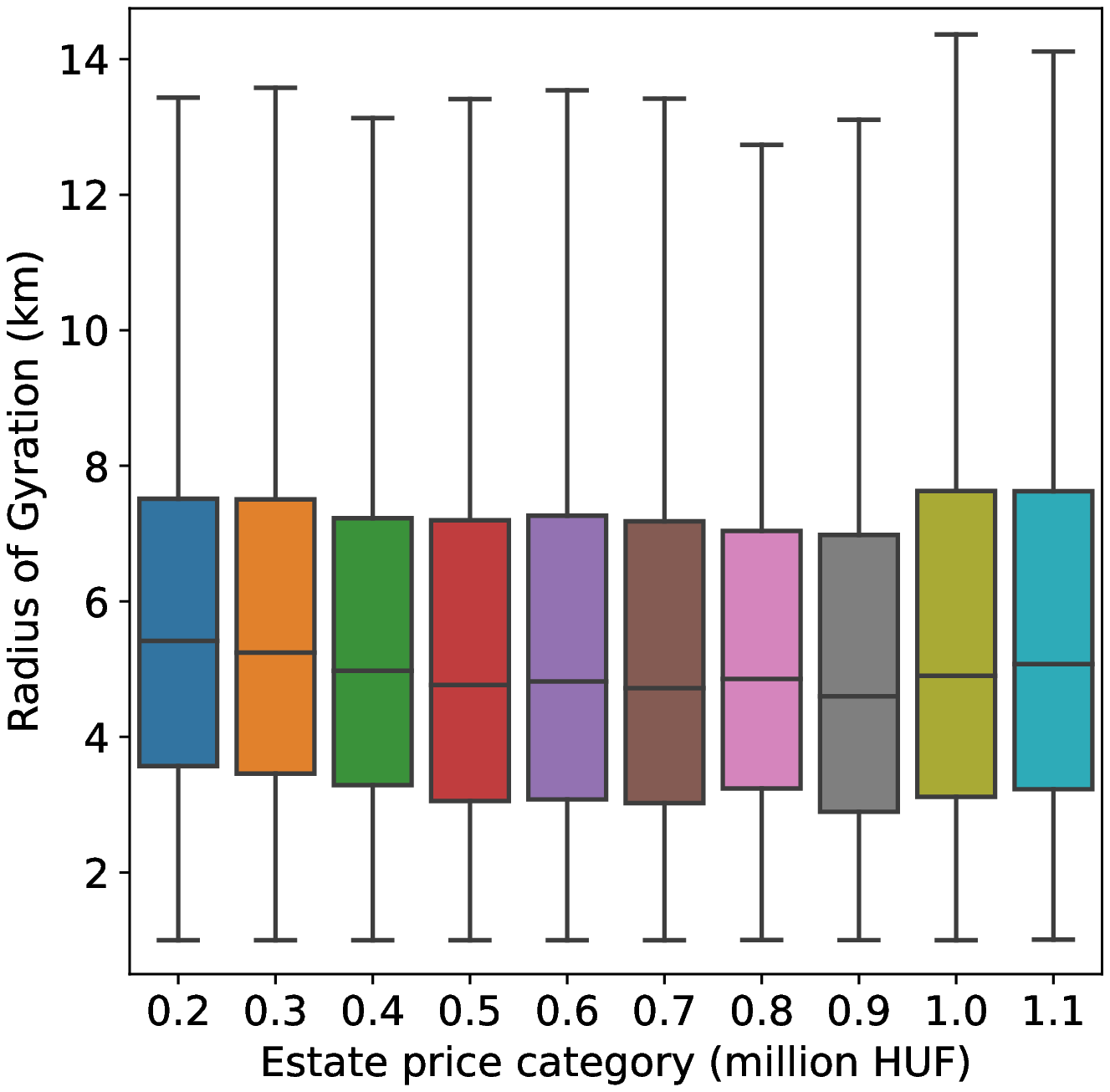}
        \caption{Radius of Gyration}
        \label{fig:gyr_pricecat}
    \end{subfigure}
    \hfill
    \begin{subfigure}[t]{0.485\linewidth}
        \centering
        \includegraphics[width=\linewidth]{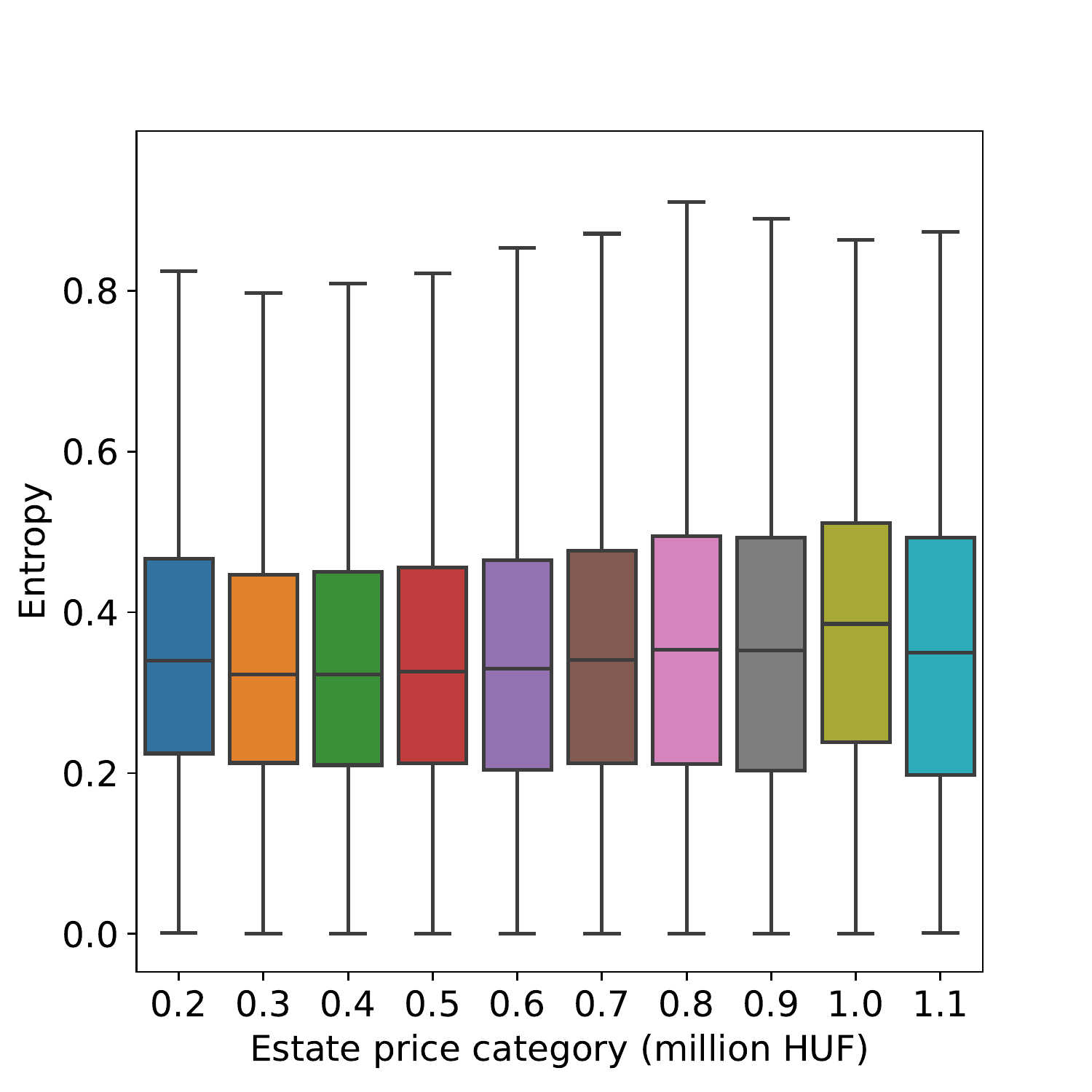}
        \caption{Entropy}
        \label{fig:ent_pricecat}
    \end{subfigure}
    \vfill
    \vspace{.5em}
    \begin{subfigure}[t]{0.485\linewidth}
        \centering
        \includegraphics[width=\linewidth]{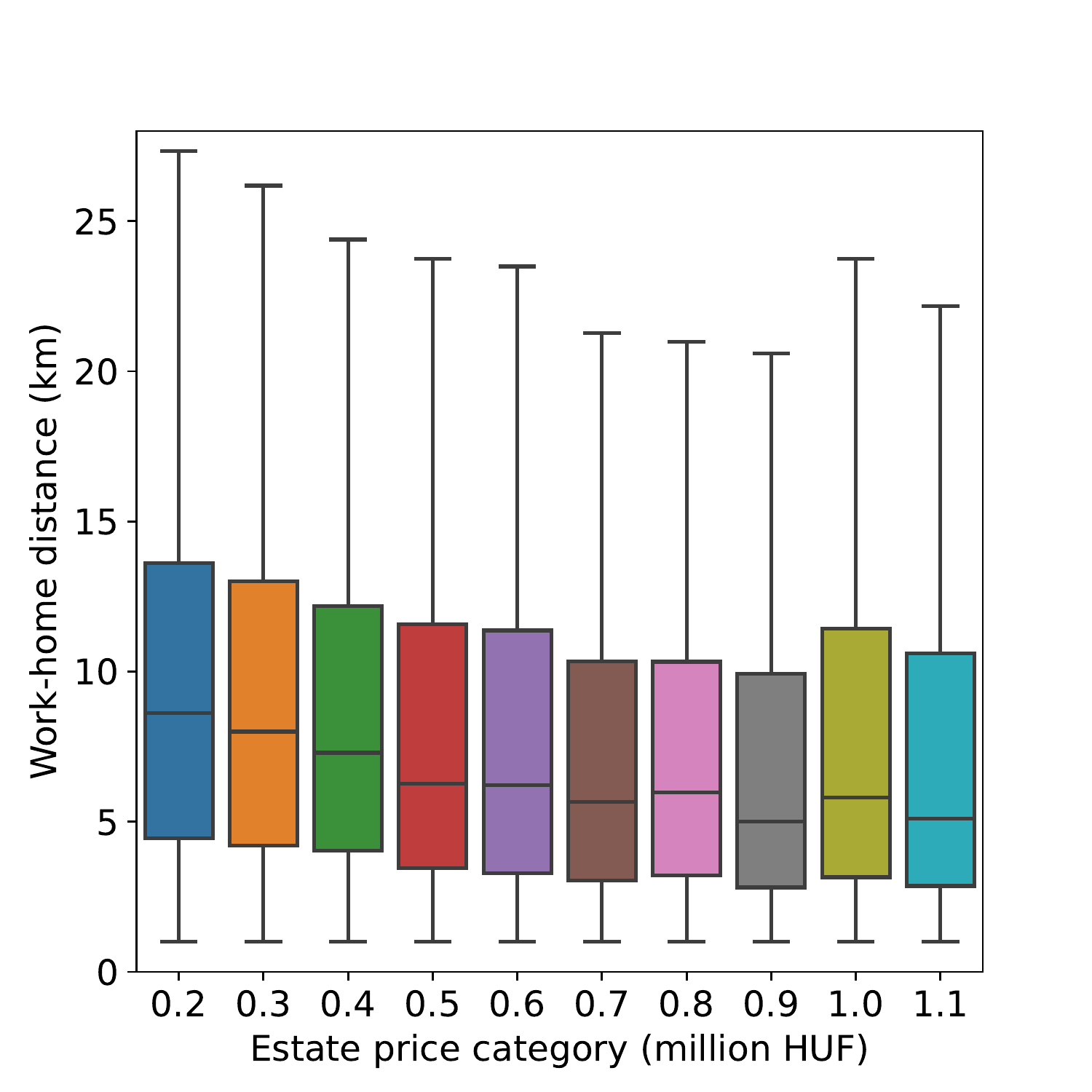}
        \caption{Home-Work distance}
        \label{fig:whd_pricecat}
    \end{subfigure}
    \hfill
    \begin{subfigure}[t]{0.485\linewidth}
        \centering
        \includegraphics[width=\linewidth]{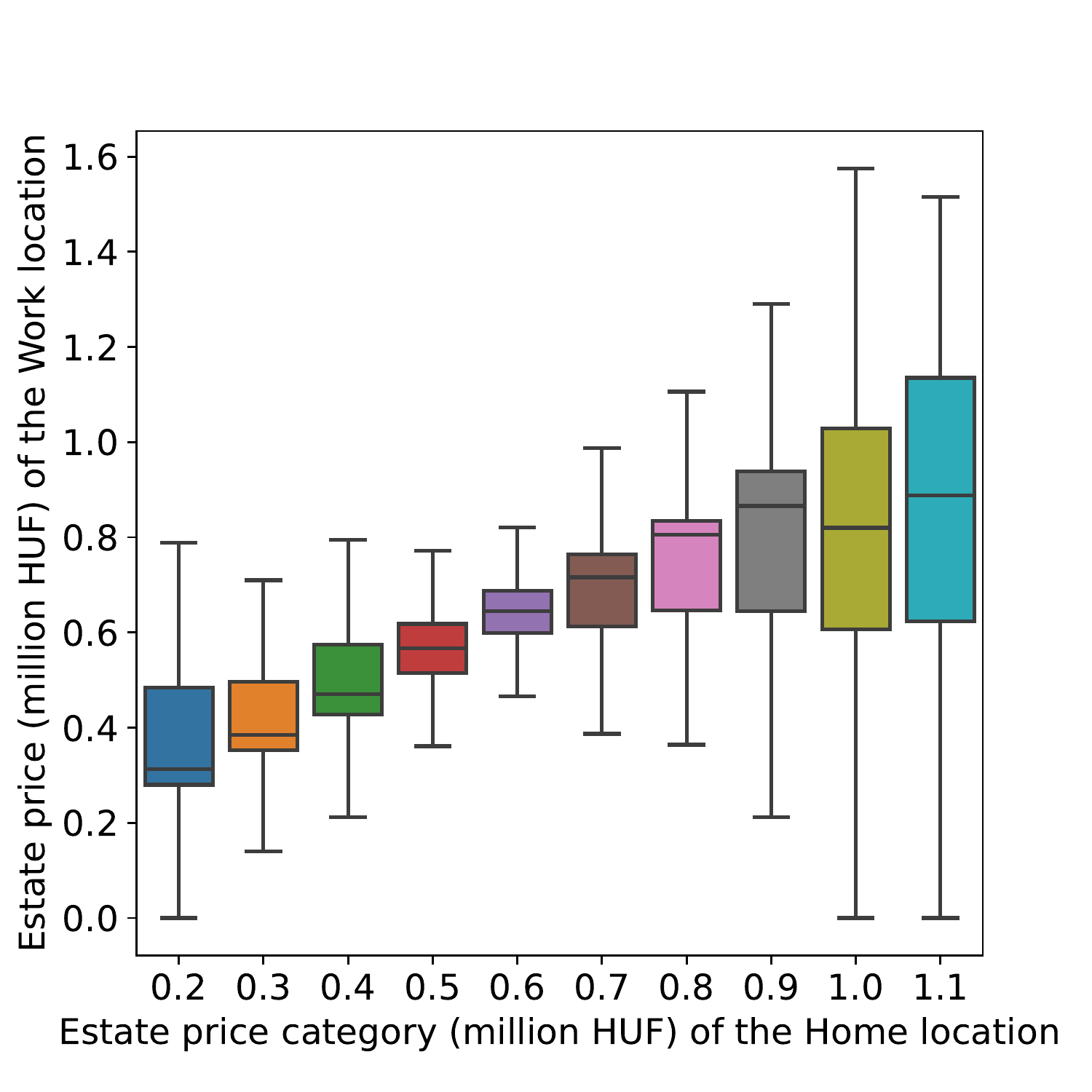}
        \caption{Estate prices of the work locations in contrast to the workers' home location estate prices}
        \label{fig:work_price}
    \end{subfigure}

    \caption{Indicators by financial category.}
    \label{fig:pricecat}
\end{figure}

The individuals are classified into ten categories, in the price range of 0.2 to 1.2 million HUF / m\textsuperscript{2}, based on the mean estate price of their home location (cell), then mean value of the indicators are determined within the classes. The Figure~\ref{fig:pricecat}, shows the Radius of Gyration, the Entropy and the Home-Work distance of the financial classes, where only the Home-Work distance (Figure~\ref{fig:whd_pricecat}) shows a slight tendency between classes.

The Radius of Gyration defines a circle in which an individual lives their life and the Entropy measures the diversity of an individual's visited locations, but both of them is highly affected by the data quality. The data used for this paper contains only those visited locations, where billed activity (phone calls, short messages or data transfer) happened, everything beyond that, is invisible. Even with only the active communication, the mobility customs of an individual can be described, but this seems to be insufficient to distinguish using financial status criteria.

The Figure~\ref{fig:work_price}, shows the typical Estate Prices of the work locations in contrast of Estate Prices of the home locations using a box plot without the outliers.
For every home location estate price category, the box plot (Figure~\ref{fig:work_price}) splits the SIM cards into three groups: (i) those between the first and the third quartile (Q1--Q3), (ii) those below the first quartile (between the minimum and the Q1: min--Q1), and (iii) those above the third quartile (between Q3 and the maximum, Q3--max). The idea was to treat the SIM card within the interquartile range (IQR) separately from those below and above the IQR.

The SIM cards are aggregated by the Home and Work location estate price categories (0.2--1.2 million HUF) and the twelve Radius of Gyration and Entropy categories (using 0.5 km distance ranges between 0.5 and 20 km for the Radius of Gyration and Entropy values with 0.05 steps between 0.05 and 1.00).
The structure of the data used for the Principal Component Analysis defined as follows.
Every row consists of 40 columns, representing 40 Radius of Gyration bins between 0.5 and 20 km and 20 columns representing 20 Entropy bins, between 0.05 and 1.00. The bins contain the number of SIM cards, that have been normalized by metrics to be able to compare them. Although the workdays and the holidays are treated separates during the whole study, the data are not explicitly labeled by them. The same table is constructed using weekend/holiday metrics and its rows are appended after the weekdays ones.
The home location estate price and the work location estate price descriptor columns are not provided to the PCA algorithm.

\section{Results and Discussion}
\label{sec:results}

In this section, the summary of the findings on the relationship of the mobility indicators and mobile phone users' social economic status is presented.

\begin{figure}[ht]
    \centering
    \begin{subfigure}[t]{0.7\linewidth}
        \centering
        \includegraphics[width=\linewidth]{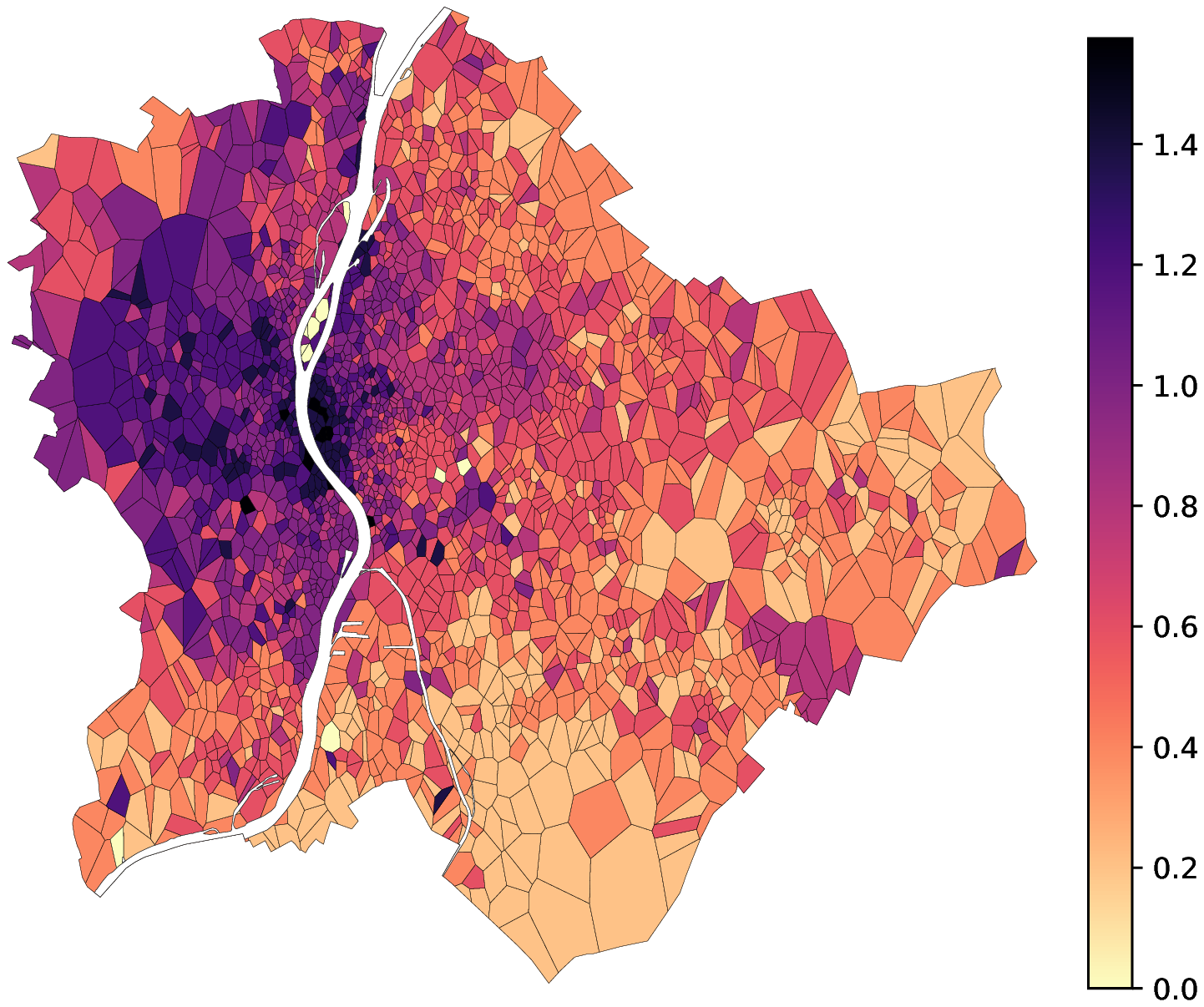}
        \caption{Cell voronoi polygons colored by estate price (million HUF)}
        \label{fig:price_map}
    \end{subfigure}
    \vfill
    \vspace{.25em}
    \begin{subfigure}[t]{0.7\linewidth}
        \centering
        \includegraphics[width=\linewidth]{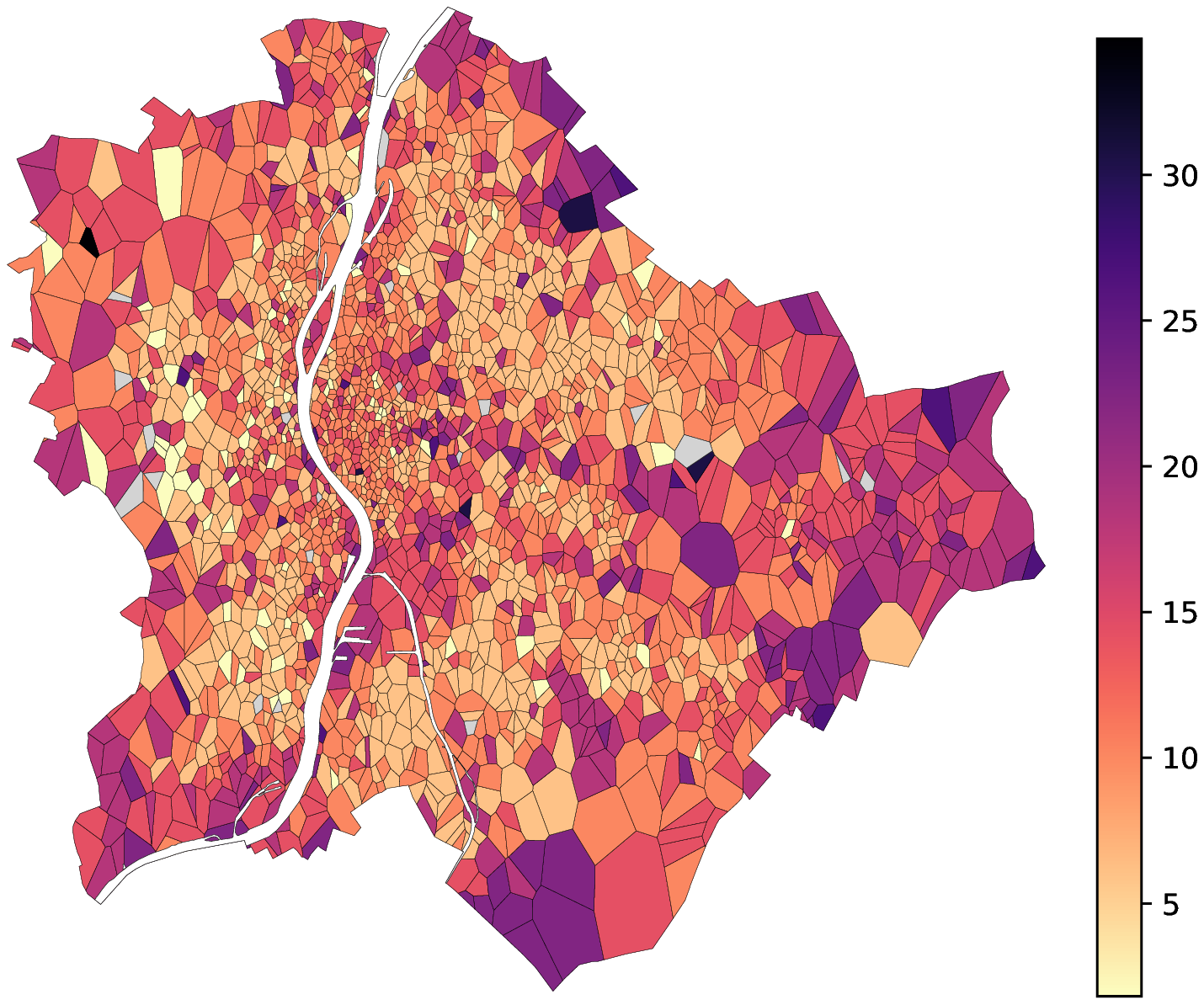}
        \caption{Cell voronoi polygons colored by Radius of Gyration (km)}
        \label{fig:gyration_map}
    \end{subfigure}
    \caption{Voronoi polygons of the merged mobile phone cells colored by the estate prices and the Radius of Gyration. River Danube is represented with white, and the cells without enough data colored with light gray.}
\end{figure}

\begin{figure}[t]
    \centering
    \begin{subfigure}[t]{0.485\linewidth}
        \centering
        \includegraphics[width=\linewidth]{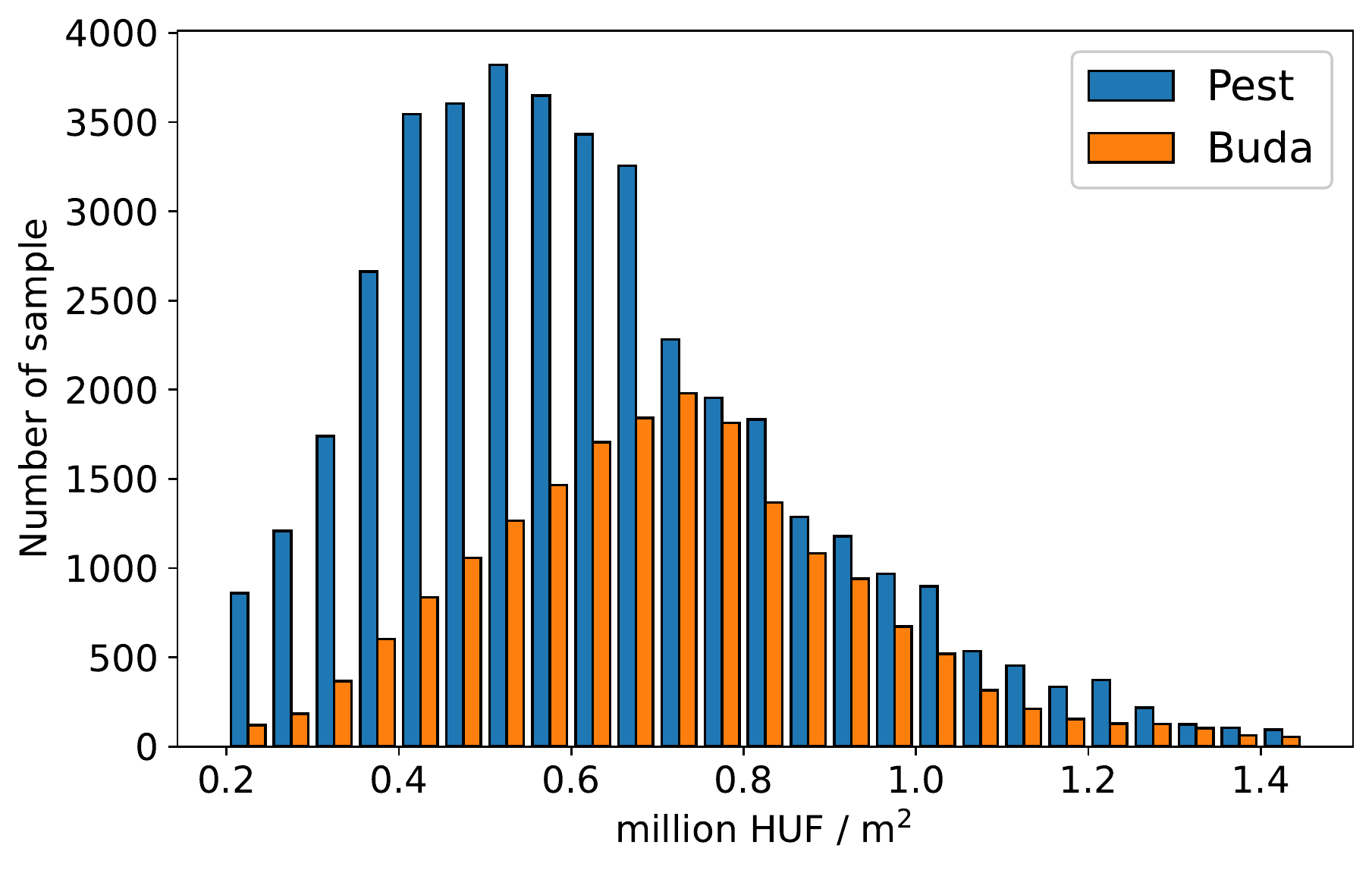}
        \caption{Distribution of the estate price samples by Buda and Pest}
        \label{fig:stacked_price}
    \end{subfigure}
    \hfill
    \begin{subfigure}[t]{0.485\linewidth}
        \centering
        \includegraphics[width=\linewidth]{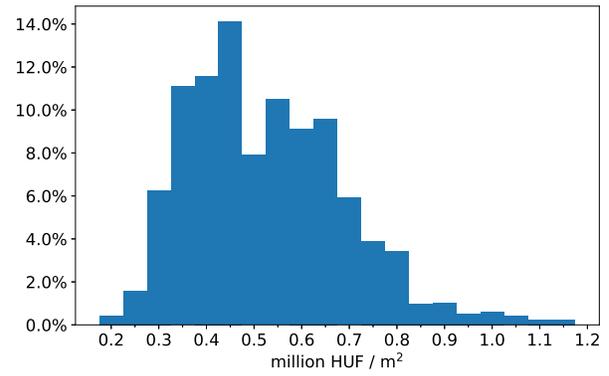}
        \caption{Normalized price distribution of the home locations (cells)}
        \label{fig:dweller_price_hist}
    \end{subfigure}

    \caption{Distributions of the estate price: (a) shows the distribution of the estate price samples distinguished by the Buda and Pest of the city, and (b) shows the distribution of the home location prices of the individuals.}
    \label{fig:norm_price_hist}
\end{figure}

Hungary is a typical capital-oriented country. Budapest is the political, economic, logistical and cultural center of the country, where almost 18\% of the population lives\cite{ksh2018budapest}. The river Danube divides the city into the Buda (Boo-da) and the Pest (Pesh-t) side, the former is the supposed more fashionable area and the property values are remarkably higher (see Figure~\ref{fig:price_map} and Figure~\ref{fig:stacked_price}).

Not every SIM works in a mobile phone and not every devices changes its position. Some device in the data set represents a 3G modem or some kind of IoT device. Though, a stationary device could also appear non-stationary in the CDR data if it changes the cell that it is connected, their movement is minimal. Thus, the devices with less than 1 km Radius of Gyration or Work-Home distance, are omitted from further analysis.

Figure~\ref{fig:price_map}, shows the spatial distribution of the estate price. The spatial distribution clearly shows that the Buda side of Budapest is much more expensive.
The cells with very light color in the middle of the city is the Margaret-Island that is a recreational area with large parks, sport establishments and hotels without any residential zone and there was no estate for sale in the data.

\begin{figure}[t]
    \centering
    \includegraphics[width=\linewidth]{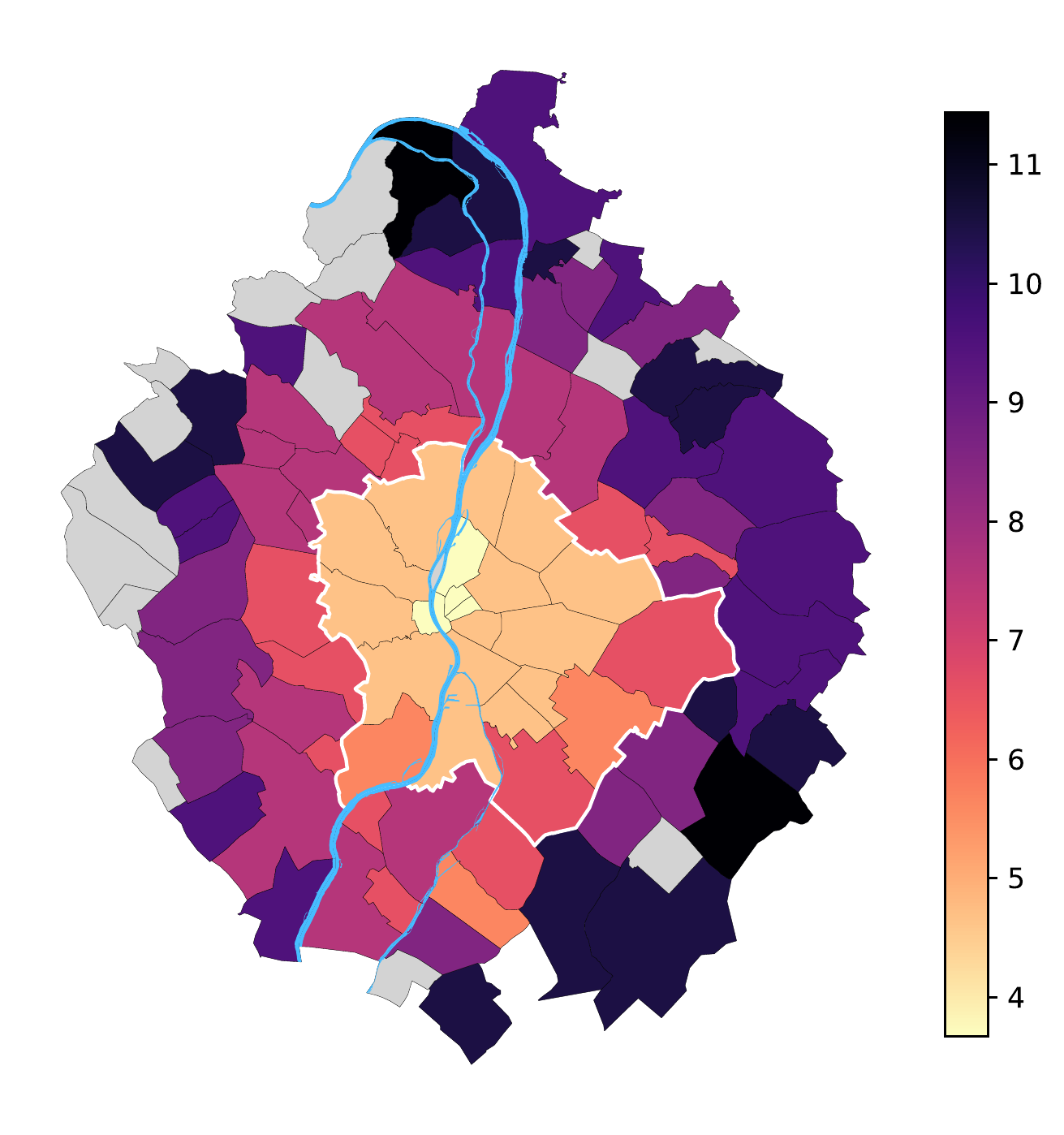}
    \caption{Budapest districts and the settlements of the agglomeration colored by the mean Radius of Gyration (km) for workdays. The tendency is that the farther lives someone from the city center the more one travels. The white border denotes the administrative border of Budapest and the Danube is displayed by light blue.}
    \label{fig:}
\end{figure}

Figure~\ref{fig:} and \ref{fig:gyration_map}, show the average Radius of Gyration of individuals whose home location is estimated to be in the given area. The former shows the Budapest suburbs and the surrounding towns and villages, the latter focuses on Budapest, at a cell level. The broad tendencies are basically the same: The farther one lives from the center, the more one travels, but the cell level map reveals some local city centers that makes the image more nuanced. The impact of the local city centers on mobility, trends could be a separate topic of investigation.

\begin{figure}[t!]
    \centering
    \begin{subfigure}[t]{0.325\linewidth}
        \centering
        \includegraphics[width=\linewidth]{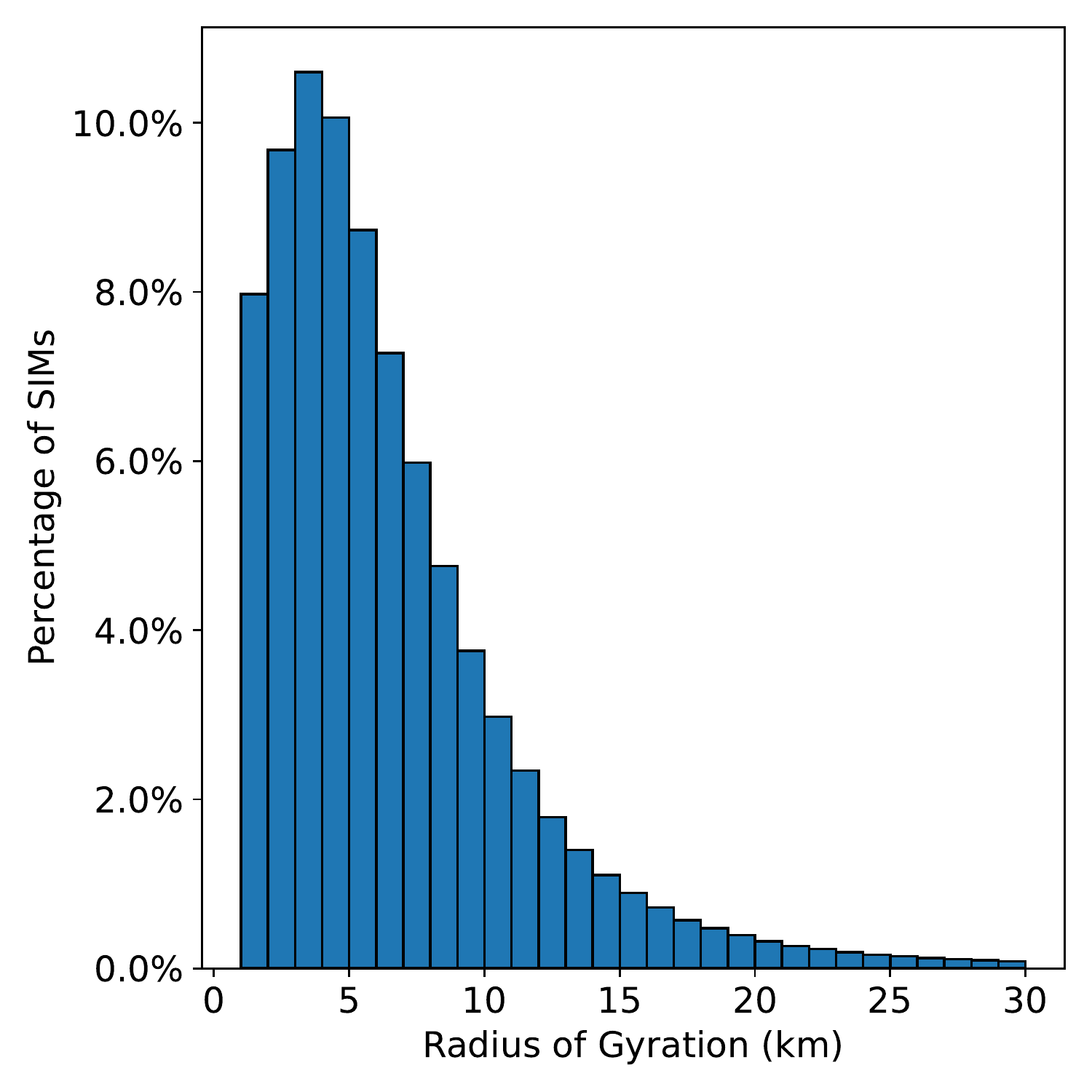}
        \caption{Radius of Gyration distribution}
        \label{fig:gyr_histogram}
    \end{subfigure}
    \hfill
    \begin{subfigure}[t]{0.325\linewidth}
        \centering
        \includegraphics[width=\linewidth]{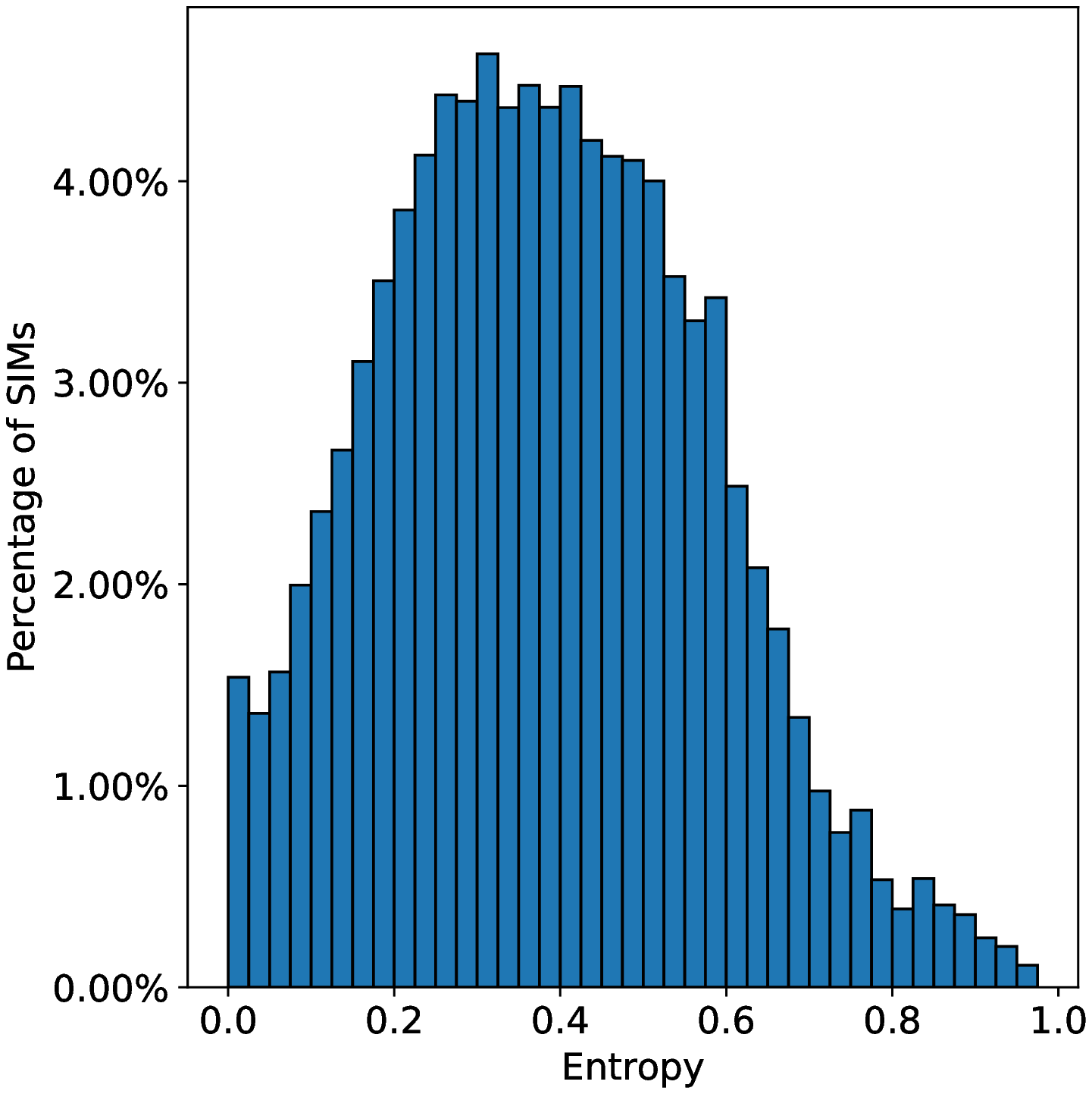}
        \caption{Entropy distribution}
        \label{fig:ent_histogram}
    \end{subfigure}
    \hfill
    \begin{subfigure}[t]{0.325\linewidth}
        \centering
        \includegraphics[width=\linewidth]{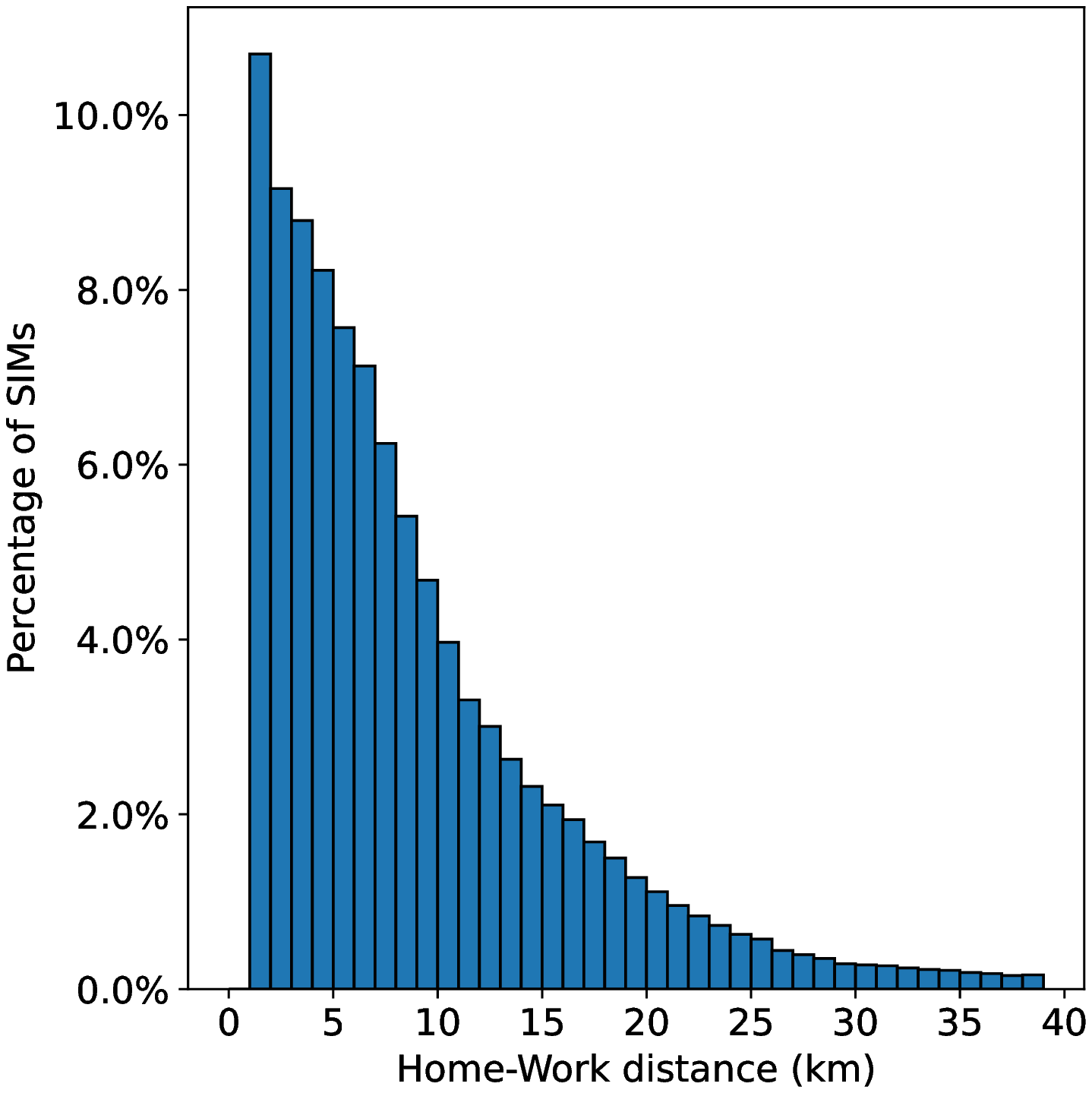}
        \caption{Home-work distance distribution}
        \label{fig:dist_histogram}
    \end{subfigure}
    \caption{Histogram of the Radius of Gyration (a), Entropy (b) and the Home-Work location distance (c)}
\end{figure}

\begin{figure}[t!]
    \centering
    \begin{subfigure}[t]{0.325\linewidth}
        \centering
        \includegraphics[width=\linewidth]{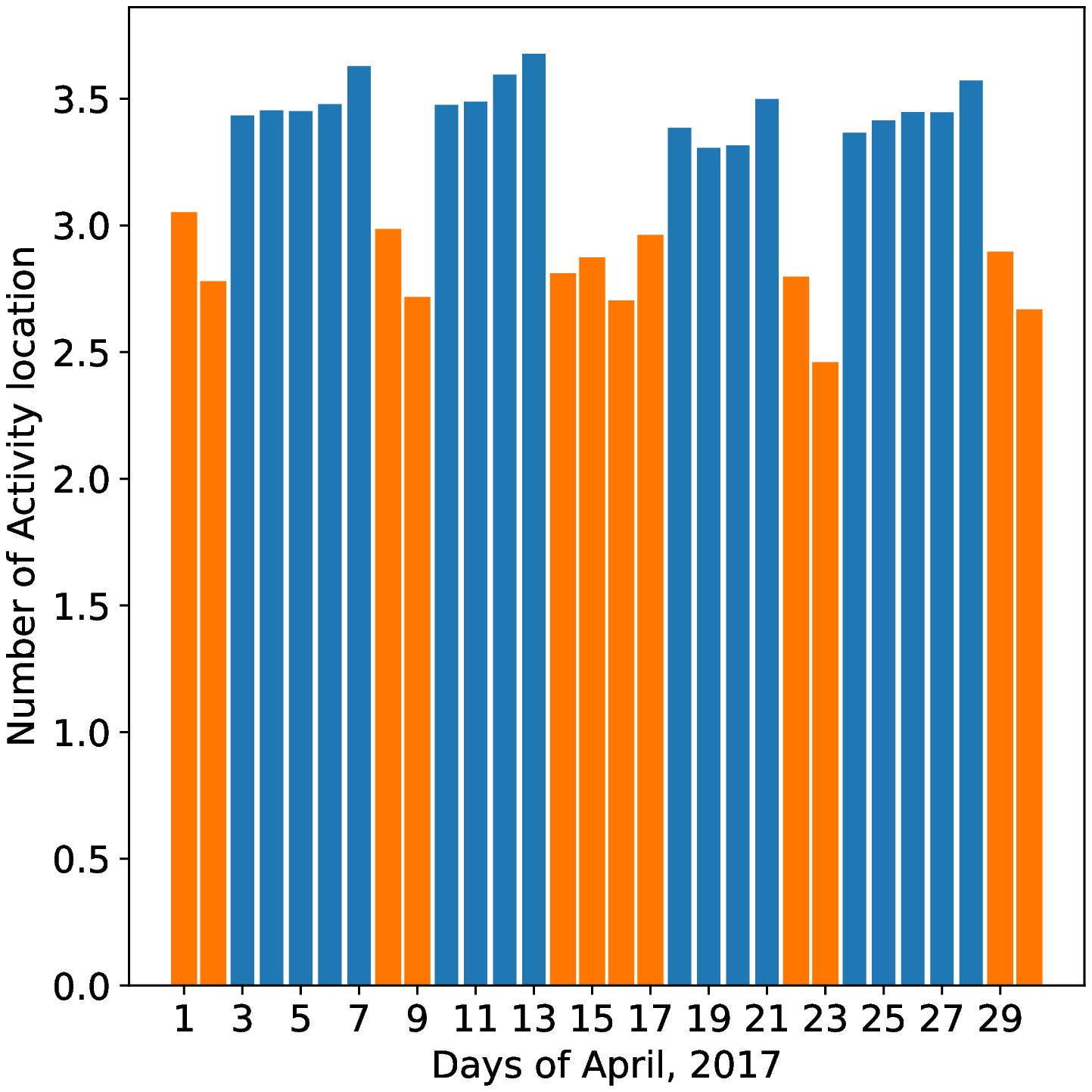}
        \caption{Radius of Gyration}
        \label{fig:gyr_daily}
    \end{subfigure}
    \hfill
    \begin{subfigure}[t]{0.325\linewidth}
        \centering
        \includegraphics[width=\linewidth]{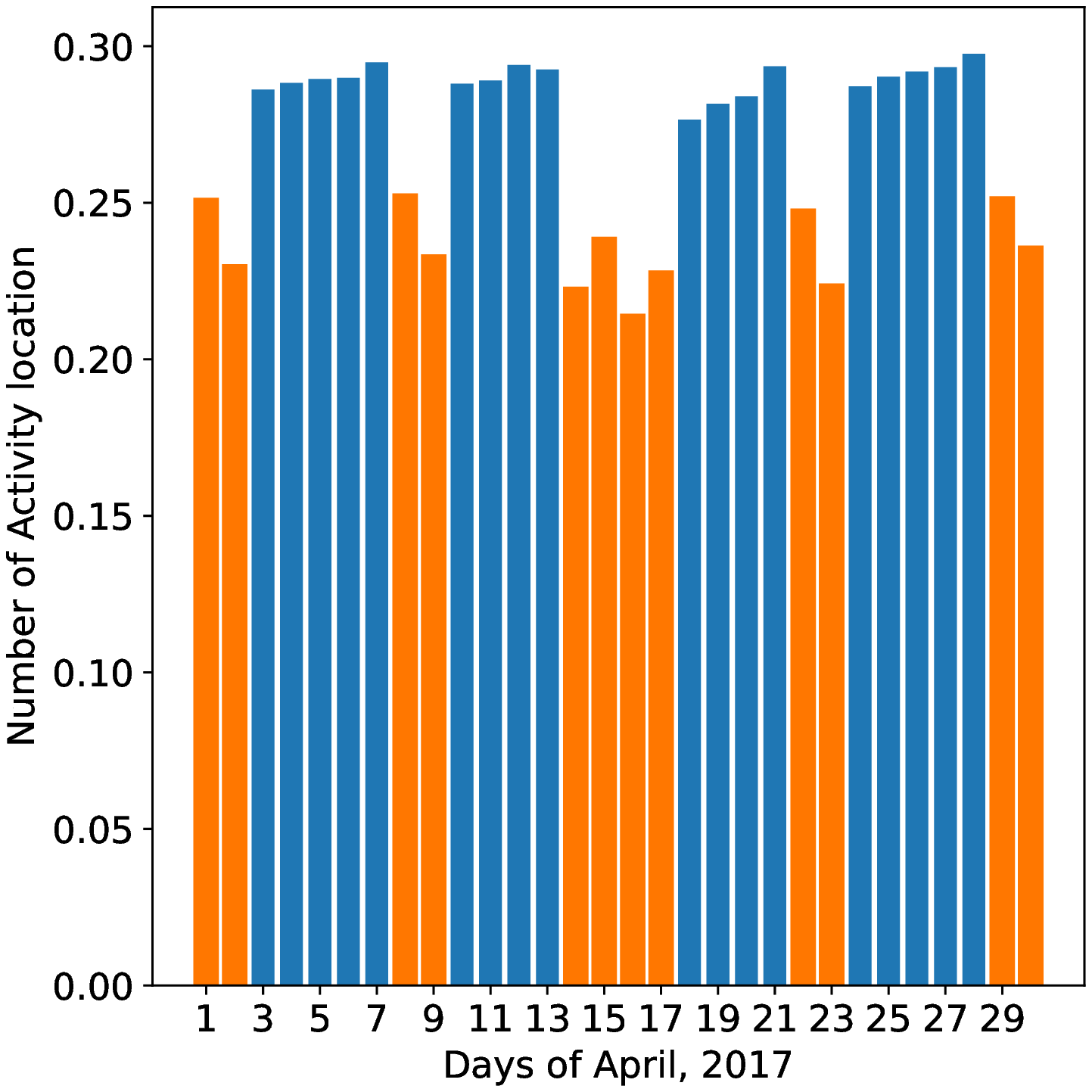}
        \caption{Entropy}
        \label{fig:ent_daily}
    \end{subfigure}
    \hfill
    \begin{subfigure}[t]{0.325\linewidth}
        \centering
        \includegraphics[width=\linewidth]{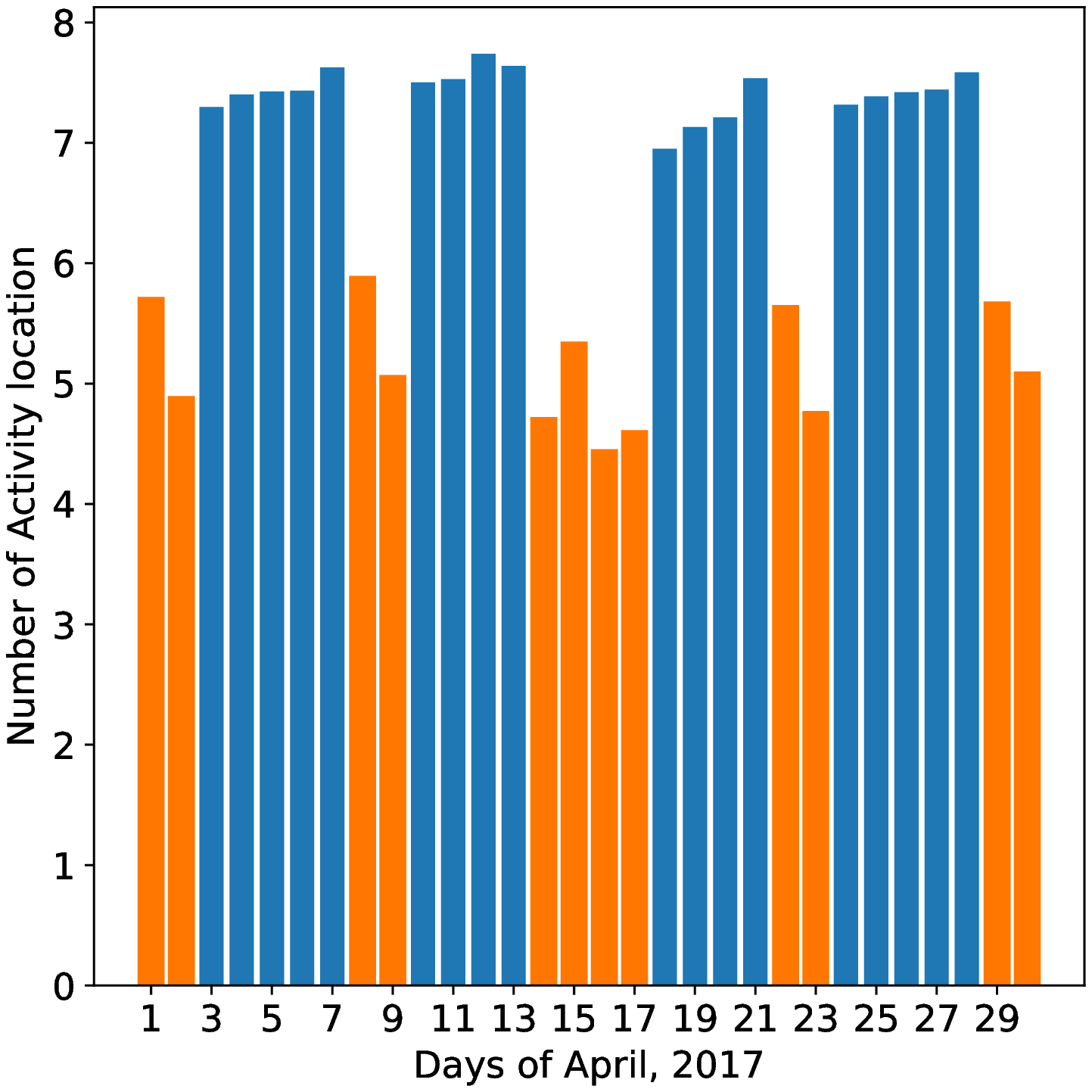}
        \caption{Number of Activity Locations}
        \label{fig:nal_daily}
    \end{subfigure}
    \caption{The differences between workdays (blue) and holidays (orange) are clear. Orange columns represent the holidays, April 14, 2017 was Good Friday and April 17, 2017 was Easter Monday that are holidays in Hungary.}
    \label{fig:daily}
\end{figure}

A social stratum model\cite{leo2016socioeconomic} has been used to differentiate the SES of users. We grouped their SES into housing prices classes. These classes represent only group-level characteristics, and cannot be used to characterize individual SES.

The mobile phone owners have been sorted in ascending order, based on their associated socioeconomic values v\textsubscript{SES}. This approach groups mobile phone users into q\textsubscript{SES} classes by the sum of socioeconomic values in each owner class is the same (i.e., same to $(\sum{v_{SES}})/q_{SES})$. The social stratum model distributes the phone users into classes with decreasing sizes (the lower the SES the higher the members in the class), and therefore the differences of SES among classes can be distinguished more effectively.

The relationships between Radius of Gyration and SES of phone users are analyzed in Budapest. We used the associated housing prices to represent the SES of the mobile phone users. The users have been ranked in ascending order based on their housing prices. The radius of gyration (Figure~\ref{fig:gyr_histogram}) shows normal distribution however it significantly skewed to the right (seems to be similar to the observations performed in Boston\cite{xu2018human}). The majority of the inhabitants have a relatively small Radius of Gyration; on the other hand, some phone users need to travel significant distances.

The Entropy shows also normal distribution. The majority of Budapest's population is visiting plenty of locations (Entropy between 0.2 and 0.6) and beside a solid fraction of people remaining the home's neighborhood (Entropy less than 0.2) there are only a few who are visiting many places in the city.

The distribution of distance between the home and work locations shows exponential characteristics Figure~\ref{fig:dist_histogram}). The majority of population investigated have chosen workplaces closer than 5 km and relatively small number of people need to travel more than 10 km to work.

The activities of the people during working days are significantly higher than on the weekends. In addition to, higher values of Gyration Radius, Entropy and the number of active locations have been recorded on the Fridays, which implies that the last working day of the week has a sort of privileged role (Figure~\ref{fig:daily}). On the weekends more activities and more vibrant travels can be observed on the Saturdays and many of the dwellers are resting on Sundays.

Figure~\ref{fig:gyr_pricecat}, represents the relationship between the radius of gyration and the normalized housing prices. We ranked the mobile phone users into 10 groups (q = 10). We found that mobile users living in middle price range (0.5-0.7 million HUF) areas have smaller Radius of Gyration on average than the cheaper (poorer) regions. Dwellers in poorer zones travels around 5 km on average while Radius of Gyration in the inhabitants from the middle class and more expensive regions is lower. This findings can be explained by the fact that mean of people living in middle price houses (Figure~\ref{fig:gyr_pricecat}) have better chance to have different job and recreation opportunities and therefore they do not need to travel longer distances. The average and the interquartile range of mobile network users living in the most expensive area (>1.0 million HUF) is larger than in the less costly area. This also follows from that fewer people have homes in more wealthy regions and therefore, the mean and the scatter of Gyration Radius is higher.

Figure~\ref{fig:ent_pricecat}, illustrates the relationships between SES and activity entropy which indicator of mobility is dedicated to describe the regularity of phone users' daily travel and activity patterns. The results show a low variability across SES for activity entropy. The mean values of each house pricing classes remained in the range of 3.5 - 4.5. One potential reason is that for most of the people, their daily activities mainly concentrate in a few locations (e.g., home and work location). However, a small increasing mean home value tendency can be observed, which suggests that the occupants living in more expensive sectors, are visiting more places. The regularity of activities, at the wealthier locations have a slight impact on the activity diversity of phone users. This finding differs from the results of the analysis\cite{xu2018human}. Xu showed that the wealth level of people, as least in Singapore and Boston, is not a limiting factor, which affects how they travel around in the city. It seems that the residents living in more expensive areas in Budapest, are more likely or forced to, visit more places than the ones having homes in the less expensive regions.

We examined the relationship between the Work-Home distances and SES. Figure \ref{fig:whd_pricecat}, illustrates the mileage between the Home and Job locations as a function of Home location prices. The descending trend mean values of distance related to home prices from the lower to the higher property prices suggest that SES has significant role on how much the dwellers have to travel to their work places. The cheapest homes are the farthest from the job location (the mean is 8.4 km and the maximum is 27.5 km), while the shortest daily work trips belong to the homes located in 0.9 million HUF category (the mean is 5.2 km and the maximum is 21.8 km). In other words, the people living in poorer area have to travel more to their work locations. More time spent on daily commuting by the people living in cheaper places which might affect their opportunities to visit more attractions in the city. This capability may also explain the relationship between activity entropy and SES (Figure~\ref{fig:ent_pricecat}).

We also analyzed correlation of property prices at home and work areas. We found that the economic status of the Home and Work locations for each mobile phone user has a directly proportional relationship (Figure~\ref{fig:work_price}). Higher mean values of housing prices, where the people spending their working hours, belong to the higher Home price classes. It is notable that the mean value of property prices in Work locations, is higher than the Home classes for residents whose housing price is between 0.2-0.6 and 0.9-1.1 million HUF. The total range (max-min) and the Q1-Q3 range of the property prices at workplaces is inversely proportional with home housing prices, in the 0.2-0.6  and directly proportional in the 0.7 - 1.1 million HUF range. This observation suggests that a small number of workplaces are in the cheaper areas of Budapest and therefore, most of the dwellers spend the working hours in the more expensive regions. On the other hand, limited workplaces are available, also in the wealthiest districts and people having home there, work in less expensive areas. The smallest scatter of the property values at workplaces, is given for the residents living in the 0.6 million HUF price regions of the city. Nevertheless, the results in Figure~\ref{fig:work_price}, suggest that job and home locations have a notable relationship.

\subsection{Characterization of SES by Principal Component Analysis}
\label{sec:pca}

The relationship between the mobility metrics (during weekdays and weekends) and the Social Economic Status (SES) is investigated using PCA.

\begin{figure}[t]
    \centering
    \includegraphics[width=\linewidth]{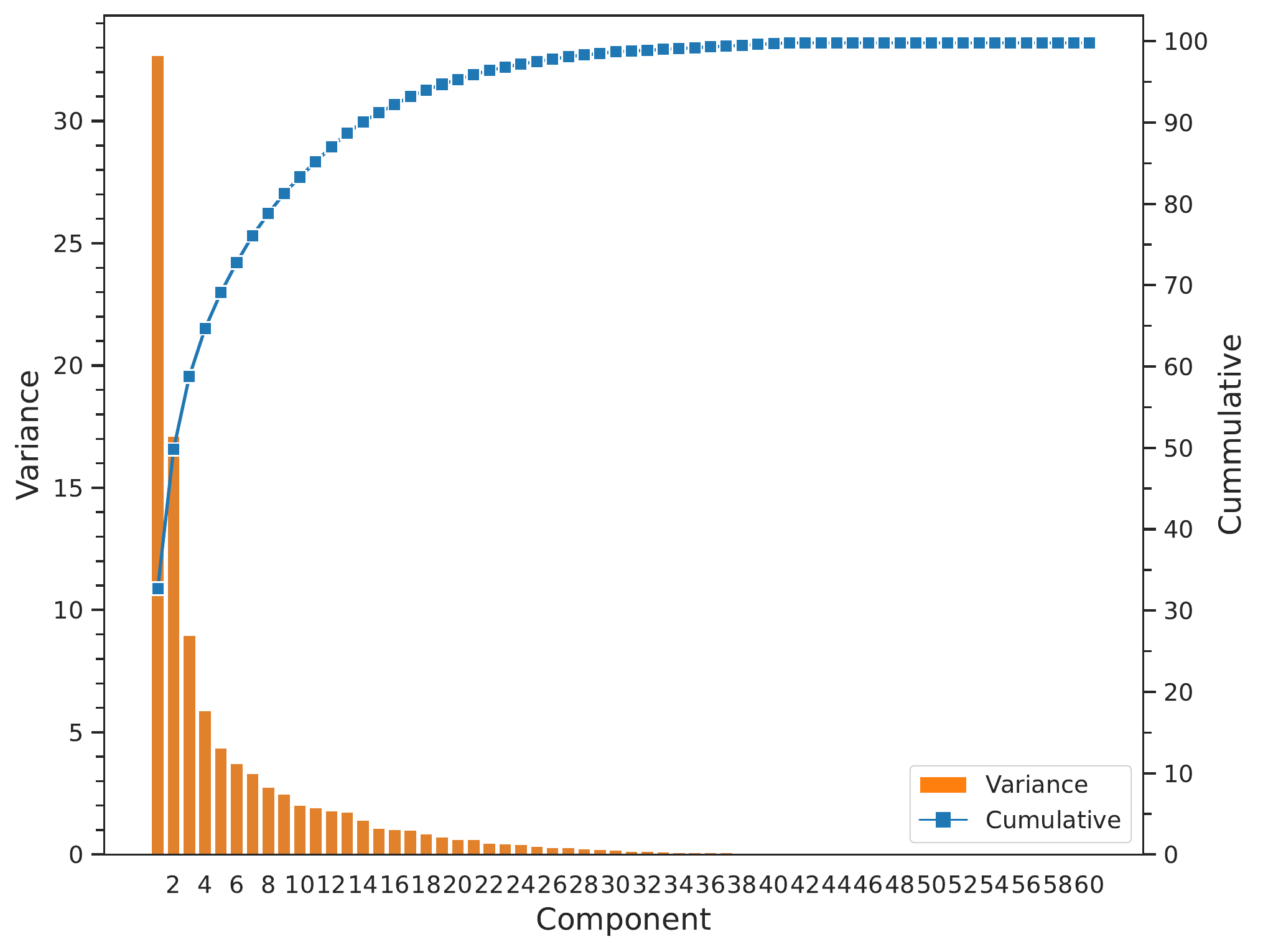}
    \caption{The Pareto histogram for the 60 components of the Principal Component Analysis}
    \label{fig:pca_var}
\end{figure}

\begin{figure}[t]
    \centering
    \includegraphics[width=\linewidth]{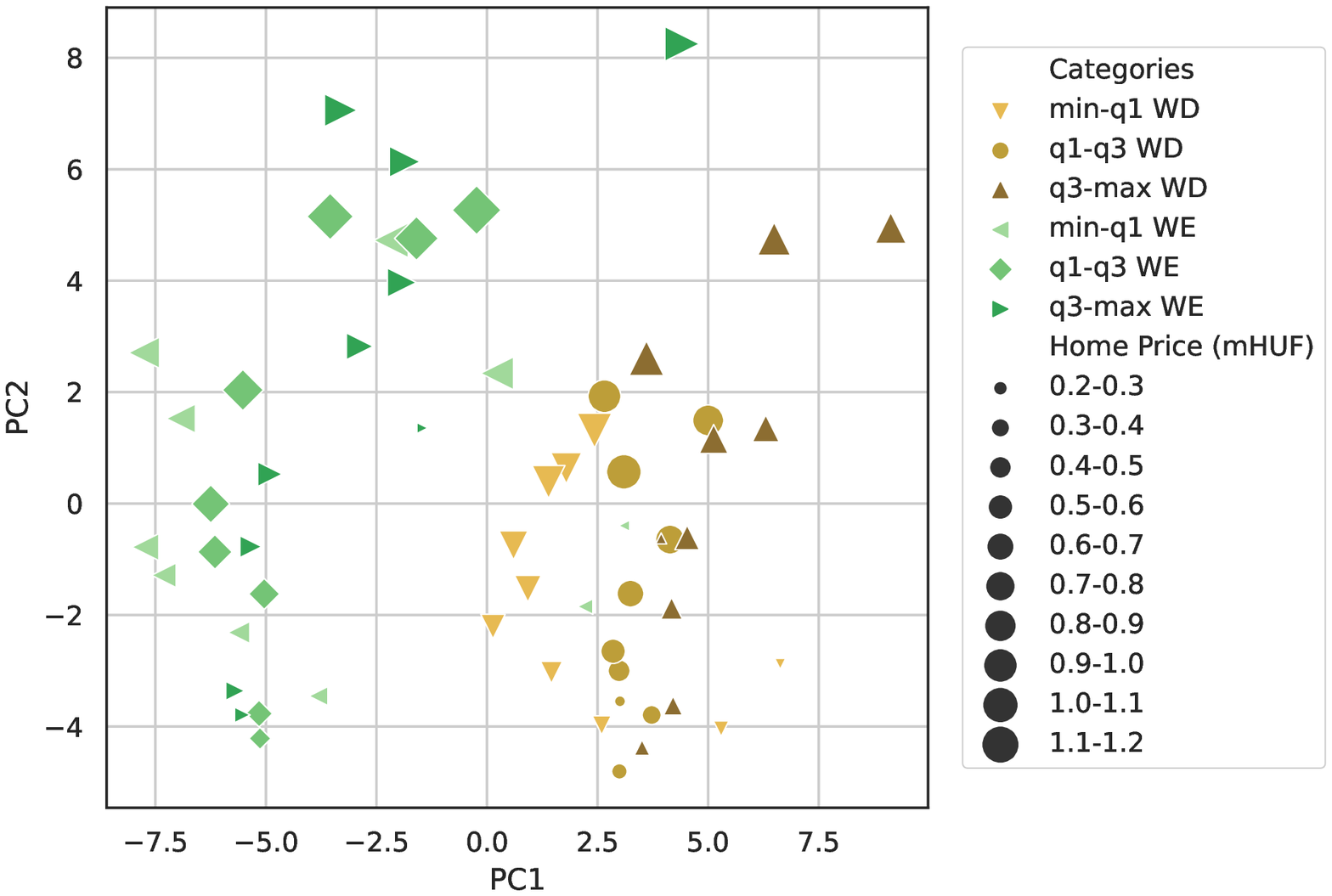}
    \caption{Scatter plot of the 2-component PCA. Marker sizes indicates the HomePrice category, the color/type WorkPrice category and also the day type (Weekday/Weekend)}
    \label{fig:pca}
\end{figure}

Figure~\ref{fig:pca_var}, shows the Pareto Histogram for the 60 components. It appears that the first two components (about 50\% of the explained variance) are sufficient to represent the variance of the variables. Figure~\ref{fig:pca}, shows the first two components of an unsupervised PCA analysis applied. The marker styles, colors and sizes each represents one label of the data.
First, the brownish colors represent the workdays, the greenish ones the holidays and they clearly form two separate clusters, though the workdays and the weekdays rows were not distinguished. This means that the first principal components are expressive enough to make a distinction between workday and holiday mobility habits.

Marker size represents the home location estate price range. There is a noteworthy tendency along the PC2 axis that the markers increase. Although the trend-line is more like a curve bending to the right, the results seems to distinguish the wealthier subscribers. The low-cost region's mobility can be found at the bottom part of the chart and the higher the property prices the higher the vertical positions of the mobility metrics. The second principal component seems to be a proper variable for representing the influence of the housing prices on gyration and entropy. It has to be noted that the smallest markers look out of place. Those represent extremely inexpensive areas, that are not residential areas, but more like industrial zones and only a minority of the subscribers are classified to those categories.
The mobility habits of dwellers living in the wealthiest regions are visibly separated from the others, especially on the working and rest days.

The mentioned quartile groups are represented with different markers and hues. The symbols representing the mobility metrics the of min-Q1 group are to the left, and the markers representing the mobility metrics of the Q3-max group are to the right of the Q1--Q3 group, in most of the house pricing groups. This means that the first principal component looks to be a determinative parameter, to distinguish not only the workday and weekend mobility patterns, but also, the quartile range of the home location estate price categories.

The range of PC1 between 0.0 and 2.5 includes markers of the first quartile for the most part, while a mixture of representatives belonging to Q1--Q3 and Q3--max groups located in the 2.5 -- 10.0 range. The weekday mobility customs of people belong to the min-Q1 group significantly differ from the other dwellers' groups. This suggests that even these residents are living in the similarly valuable neighborhood like the ones of the groups Q1--Q3, they are managing their lives differently. It is notable that the deviation is smaller in the lowest and the highest housing price categories while bigger in the middle level property price range. It seems that people having homes in the middle price level area in the city having more diverse habit than the ones living in the poorer or wealthier districts.

The characteristics of mobility, related to the dwellers associated with the Q1--Q3 and Q3--max groups, are quite similar. In the range of PC1 between 2.5 and 5.0, several markers from different groups and different housing price categories overlap each others. The markers on the chart (i.e., in the range of PC2 between 0.0 and -0.2) probably show that residents living in cheapest and the middle price level categories having very similar traveling patterns independently if they belong to the Q1--Q3 and Q3--max groups. It is also remarkable, that during the weekdays and the weekends, mobility customs of the most wealthiest districts' population are completely separated from other's. This observation suggests that the daily traveling routines of the residents having homes at the most expensive area of Budapest are significantly different than the others.

\subsection{Limitations of this Study}

There were a few limitations of this study. We associated inhabitants with the mean property value at their home cell location (i.e., Voronoi polygons). In some regions of the city, where the density of the cellphone towers is low, this approach could led to inaccuracy on the SES of residents (the housing prices vary on relatively bigger zones).
Only one variable (housing price) has been used, other socio-demographic factors are not taken into consideration. Family and household size, education level, etc. are essential dimensions of SES.
Future studies are planned to involve more parameters describing inhabitants economic characteristics.

\subsection{Future Work}

One possible research enhancement would be to use more precise data sources for the financial classification, for example, the type and price of mobile phones. The combination of the estate price around home location and the price of the mobile phone device might provide a better result.

Using a data set that contains information about both participants of a call, not only the social network could be built, but that network could be weighted with the financial status of the subscribers.

It may be possible to enhance of this work by examining the visited places, of the wealthiest classes, whether they tend to appear in areas with more elegant restaurants, theaters, etc. outside of normal work hours.

\section{Conclusions}
\label{sec:conclusions}

In this study, the travel customs of a large population in Budapest (Hungary) has been investigated using Call Detail Records. Along with mobility indicators, computed from the mobile phone network data, real estate price is used to represent the socioeconomic status of the people living in a given area.
In order to find and evaluate the correlation between human mobility patterns and an inhabitants socioeconomic status, in Budapest, an analytical framework has been applied, on the coupled datasets of mobile phone networks and real estate prices.

The Radius of Gyration, Entropy and Euclidean distance between Home and Work locations has been derived from the mobile phone network data and used as mobility indicators to characterize the travel patterns of inhabitants. The property price of homes has been applied to represent their socioeconomic status. We performed data fusion method to quantify at an aggregate level the SES (Socioeconomic status) of mobile phone users.

By investigating the relationships between SES and three mobility indicators, we found that wealth level in Budapest has a certain influence on travel customs. The real estate price of home sites, does not have remarkable effect on the number of visited places (Entropy) and Gyration, which suggests that the travel diversity of the inhabitants, is not strongly dependent from their socioeconomic status.

On the other hand, the mean distance of home and work locations are significantly correlated to property values. The people living cheaper regions have to take longer routes to their work sites. The reason for this relationship, can be explained by the fact that the less expensive districts are located at the perimeter of the city, while most of the work sites are in the wealthier regions. Analyzing the real estate values at home and work locations of the residents, we found a strong positive relationship. Living in a high-priced neighborhood in Budapest, requires working in more expensive regions and vice versa.

The Principal Component Analysis revealed similarities and differences of movement behavior of residents living in different price level categories. The correlations of mobility customs and SES found could be used efficiently for supporting the activities of public transportation, real estate developments and retail.

We also evaluated the connection of movement behavior and SES by performing PCA (Principal Component Analysis). The results showed that the first two components could be sufficient to distinguish the workday and weekend mobility patterns, as well as the home location estate price categories.
The mobile phone network data can also be used to analyze commuting, similar to census data, even though its spatial resolution is not as accurate. The commuting trends between the given sectors of the agglomeration and the district groups of Budapest, match enough to say, that Call Detail Records can be a cheaper and temporally, more accurate, solution for commuting and possibly other kinds of sociological studies.

In conclusion, this research focused on identifying hidden or unclear links, between human mobility and socioeconomic status. To that end, this work has been successful. The approach and even the framework developed herein, can be applied for various locations. The findings can be used to understand more precisely and efficiently, the conditions and circumstances, of day-to-day life, for people, living in our modern urban environments.

\vspace{6pt}

\section*{Author Contributions}
Conceptualization, G.P. and I.F.; methodology, G.P.; software, G.P.; validation, G.P. and I.F.; formal analysis, G.P.; investigation, I.F.; resources, I.F.; data curation, G.P.; writing---original draft preparation, G.P. and I.F.; writing---review and editing, G.P. and I.F.; visualization, G.P.; supervision, I.F.; project administration, I.F.; funding acquisition, I.F. All authors have read and agreed to the published version of the manuscript.

\section*{Funding}
This research supported by the project 2019-1.3.1-KK-2019-00007 and by the Eötvös Loránd Research Network Secretariat under grant agreement no. ELKH KÖ-40/2020.

%
%

\section*{Acknowledgments}
The authors would like to thank Vodafone Hungary for providing the CDR data.
Map data copyrighted by the OpenStreetMap contributors, and licensed under the Open Data Commons Open Database License (ODbL).
The estate price data are provided by the \url{ingatlan.com} estate selling portal.

\section*{Conflicts of Interest}
The authors declare no conflict of interest. The funders had no role in the design of the study; in the collection, analyses, or interpretation of data; in the writing of the manuscript, or in the decision to publish the results.

%

\printbibliography

\end{document}